\documentclass[aps,prd,preprint,superscriptaddress,nofootinbib]{revtex4}

\usepackage{amsmath} \usepackage{amsfonts} \usepackage{amssymb}
\usepackage{wasysym}
\usepackage[dvipsnames]{xcolor}
\usepackage{graphicx}
\usepackage{multirow}
\usepackage{here}
\usepackage{epsfig}
\usepackage{epstopdf}
\usepackage{soul}
\usepackage{hyperref}
\usepackage{float}
\usepackage{xcolor}
\usepackage{graphicx}
\usepackage{multirow}
\usepackage{ulem}
\usepackage{array}

\newcolumntype{L}{>{\displaystyle}l}
\newcolumntype{C}{>{\displaystyle}c}
\newcolumntype{R}{>{\displaystyle}r}

\newcommand{\fn}[2]{\mathinner{#1\mathopen{\left(#2\right)}}}

\newcommand{\sref}[1]{Section~\ref{#1}}

\newcommand{\be}{\begin{equation}}
\newcommand{\ee}{\end{equation}}
\newcommand{\bea}{\begin{eqnarray}}
\newcommand{\eea}{\end{eqnarray}}

\begin{document}

\title{Trapped Gravitational Waves in Jackiw-Teitelboim Gravity}

\author{Jeong-Myeong Bae}
\email{bjmhk2@dgist.ac.kr}
\affiliation{School of Undergraduate Studies,
College of Transdisciplinary Studies, DGIST,
Daegu 42988, Republic of Korea}

\author{Ido Ben-dayan}
\email{ido.bendayan@gmail.com}
\affiliation{Physics Department,
Ariel University, Ariel 40700, Israel}

\author{ Marcelo Schiffer}
\email{schiffer@ariel.ac.il}
\affiliation{Physics Department,
Ariel University, Ariel 40700, Israel}

\author{Gibum Yun}
\email{kbyun97@dgist.ac.kr}
\affiliation{School of Undergraduate Studies,
College of Transdisciplinary Studies, DGIST,
Daegu 42988, Republic of Korea}

\author{Heeseung Zoe}
\email{heezoe@dgist.ac.kr}
\affiliation{School of Undergraduate Studies,
College of Transdisciplinary Studies, DGIST,
Daegu 42988, Republic of Korea}

\date{\today}

\begin{abstract}
We discuss the possibility that gravitational fluctuations ("gravitational-waves")  are trapped in space by gravitational interactions in two dimensional Jackiw-Teitelboim gravity. In the standard geon (gravitational electromagnetic entity) approach, the effective energy is entirely deposited in a thin layer, the active region, that achieves spatial self-confinement and raises doubts about the geon's stability. In this paper we relinquish the "active region" approach and obtain self-confinement of  ``gravitational waves'' that are trapped by the vacuum geometry and can be stable against the backreaction due to metric fluctuations.
\end{abstract}

\maketitle

\section{Introduction}

In 1916, Einstein predicted that gravitational sources could produce waves of spacetime from his theory of general relativity \cite{Einstein:1916cc}. 
In 1955, Wheeler introduced a particle-like object, geon (gravitational electromagnetic entity), where gravitational perturbations are confined in space because of electromagnetic interaction \cite{Wheeler:1955zz}. 
He hoped to construct the geon as an elementary particle but that did not seem fruitful. Brill and Hartle elaborated this idea by considering gravitational waves (GW) trapped by gravitational interactions \cite{Brill:1964zz}, i.e., that GW are somewhat localized in space by their self-interaction. Given the dispersive nature of radiation, it seems such objects are metastable at best. Analyses in general relativity have devoted much effort to the discussion of whether such a solution is self-consistent and metastable \cite{Cooperstock:1995cq, Cooperstock:1995yv, Anderson:1996yb, Anderson:1996pu,Perry:1998hh}. These analyses assumed an empty asymptotic Minkowski background.  Needless to stress the importance of considering  stable self-confining gravitational configurations having as  the background geometry the  Friedman-Lemaitre--Robertson--Walker (FLRW) Universe or at least, asymptotically de Sitter (dS). Significant works have been done on asymptotic Anti-de Sitter in \cite{Martinon:2017esl} and references therein. 

 In this paper, we study fluctuations of the gravitational field (``gravitational waves'') trapped in space by the vacuum geometry in the framework of classical two-dimensional (2D) Jackiw--Teitelboim (JT) gravity \cite{Teitelboim:1983ux, Jackiw:1984je}.   
We prefer to use the term ``trapped gravitational waves'' instead of ``geon'' because in the classical geon solution the effective energy-momentum that corrects the unperturbed solution is entirely deposited in a thin shell enclosing the geon (active region). Our motivation is to point towards a different kind of self-gravitating clump, a different paradigm that circumvents the need for an active region. Clearly, the choice of 2D gravity stems from the tremendous simplification of calculations. However, in 2D, the Einstein tensor vanishes identically and Einstein's equations are trivial. We therefore choose JT gravity as an alternative for a simple gravity theory in 2D that has a cosmological constant (CC) and dynamical solutions.

Generally speaking, in the vacuum, there is a competition between the gravitational perturbations that travel at the speed of light and disperse, and their self-gravitational pull. The motivation of a non-vanishing CC comes from the intuitive fact that it further generates a potential such that our solution sits in the vicinity of the potential minimum. This result corresponds to trapped ``gravitational waves''.
To fully understand the structure of the theory, we thoroughly discuss different gauges and independent degrees of freedom in the theory. More specifically, we study perturbations in the traceful gauge that is volume changing, and perturbations in traceless gauges, that better mimic GW gauges. Our analysis yields that perturbations can be trapped in some region of space. 
Furthermore, we discuss possible gauge issues, the connection between solutions in various coordinate systems and provide several examples. 
 
The paper is organized as follows.
In \sref{geon_vacuum}, we apply the method of \cite{Anderson:1996pu} for finding gravitational geons to JT gravity. 
In \sref{gauge} we discuss possible gauge transformations and what degrees of freedom remain after using up the gauge freedom.
In \sref{GW},  and \sref{Marcelo1} we find analytic and numerical trapped solutions in various gauges.  
In \sref{synchronous}, we display the exact solution in the synchronous, conformal and spatially flat frame of references that exhibit a wave behavior and sketch similar trapped solutions.
\mbox{In \sref{dis}, }we summarize our results and discuss future directions. 
  
\section{Finding a Geon in Jackiw-Teitelboim Gravity}\label{geon_vacuum}

Our starting point is 2D gravity introduced by Jackiw and Teitelboim \cite{Teitelboim:1983ux, Jackiw:1984je}, where the equation of motion is given by 
\begin{equation}
\mathcal{R} - \Lambda = 8 \pi G T,
\label{eq:Einstein}
\end{equation}
where $\mathcal{R}$ is the curvature scalar, $\Lambda$ is the CC, $T$ is the energy-momentum and $c=1$. Notice that in 2D Newton's constant is dimensionless and can always be absorbed into the gravitational field. 
As in \citep{Sikkema:1989ib, Vollick:2008ha}, we take the following metric ansatz: 
\begin{equation}
g_{a b} = \gamma_{ab} + h_{ab},
\end{equation}
where $\gamma_{ab}$ is the unperturbed metric with signature $(-,+)$ and $h_{ab}$ represents the perturbations. 

If we consider no matter, i.e., $T=0$,  Equation (\ref{eq:Einstein}) becomes
\begin{equation}
\fn{\mathcal{R}}{\gamma_{ab},h_{ab}} = \Lambda
\end{equation}

Following 
 \citep{Brill:1964zz, Anderson:1996pu}, we expand it perturbatively as 
\begin{equation}
\fn{\mathcal{R}^{(0)}}{\gamma_{ab}} + \fn{\mathcal{R}^{(1)}}{\gamma_{ab},h_{ab}} + \fn{\mathcal{R}^{(2)}}{\gamma_{ab},h_{ab}} \simeq \Lambda
\end{equation}
where $(0), (1), (2), \ldots$ imply the orders in $|h_{ab}|\ll 1$. 
We then solve this equation in three steps:
First, the background geometry for the vacuum state comes from 
\begin{equation}
\fn{\mathcal{R}^{(0)}}{\gamma_{ab}} = \Lambda~.
\label{eq:Rzeroth}
\end{equation}

Second, the first order perturbation equation in $h_{ab}$
\begin{equation}
\fn{\mathcal{R}^{(1)}}{\gamma_{ab},h_{ab}} = 0~,
\label{eq:Rfirst}
\end{equation}
is a wave-type equation. 
 Hence, the gravitational waves $h_{ab}$ trapped in space are determined by   
\eqref{eq:Rfirst}.
Third, we test the stability of the solution by considering the backreaction on the metric through
\begin{equation}
\fn{\mathcal{R}^{(0)}}{\tilde{\gamma_{ab}}} + \langle \fn{\mathcal{R}^{(2)}}{\tilde{\gamma_{ab}},h_{ab}}  \rangle  = \Lambda 
\label{eq:Rsecond}
\end{equation}
where the original metric $\gamma_{ab}$ changes into $\tilde{\gamma}_{ab}$ due to the backreaction and $\langle \cdots \rangle $ means time average.

\section{Extraction of Physical Degrees of Freedom}\label{gauge}
When considering perturbations off some background metric, it is important to properly count the correct number of physical degrees of freedom that should be gauge invariant. 
Considering the general perturbed metric $g_{ab}=\gamma_{ab}+h_{ab}$, the wave Equation \eqref{eq:Rfirst} can be written as:
\be 	
\label{eq:R1general}
\fn{\mathcal{R}^{(1)}}{\gamma_{ab},h_{ab}} =h^{ab}_{\quad;ab}-\Box \tilde h-\frac{1}{2} \tilde h \mathcal{R}^{(0)}(\gamma_{ab})=0,
\ee
for any choice of coordinate system or gauge, where a semicolon denotes a covariant derivative, $\square$ is the covariant D'Alambertian, and $\tilde h\equiv \gamma^{ab}h_{ab}$ is the trace. 
This expression seems to suggest that for the purpose of calculations there are two preferred gauges, traceless and Lorentz. This is a proper time for pausing the calculations and discussing the gauge freedom in 2D.
Coordinate transformations can be represented by gauge transformations
\begin{equation}
{h}'_{ab}={h}_{ab} -\xi_{a;b}-\xi_{b;a}
\end{equation}
for any vector  $\xi^a$, being of the same order of magnitude of $h_{ab}$ itself. As a side remark, solving the field equations for
\begin{equation}
	\overline{h}'_{ab}={h}_{ab} -\frac{1}{2}g_{ab} \tilde h
\end{equation}
as it is usually done in 4D, is pointless as this relation cannot be inverted to obtain $h_{ab} $ since  in 2D the trace  $\overline{h}\equiv \overline{h}_a^a$ vanishes identically. This is somewhat reminiscent of the fact that the Einstein tensor is trivial in 2D. Consequently, all the discussion of gauge invariance must be done in terms of $h_{ab} $.
In 2D we can always express any vector as
\begin{equation} \label{eq:xiv}
	\xi_a=\phi_{,a}+\epsilon_{ab}\psi^{,b}
\end{equation}
for two different scalar fields  $\phi ,\psi $ and  $\epsilon_{ab} $ stands for the Levi--Civitta in 2D 
\begin{equation}
	 \epsilon_{ab}=\sqrt{(-g)} [a,b],
\end{equation}
where
\begin{equation}
	[0,1]\equiv1\quad ; \quad [0,0]=[1,1]\equiv0\quad; \quad [1,0]\equiv-1
\end{equation}

\subsection{Lorentz Gauge}
It is always possible to make a gauge transformation which brings a general perturbation $h_{ab}$ to the Lorentz gauge. Consider the divergence of a desired gauge transformation,
 \begin{equation}
 	h^{'b}_{\,\,\,a;b}=h_{a\,\,;b}^{\,\,\,b} -
\xi_{a\quad;b}^{\quad ;b}-\xi^b_{;\,\,ab}=0.
 \end{equation}
The commutation of covariant derivatives satisfies 
\begin{equation}
\xi^b_{;ba}-
\xi^b_{;ab}=-\mathcal{R}_{ad} \xi^d.
\end{equation}
Furthermore, recall that in 2D 
\begin{equation}
	\mathcal{R}_{ab}=\frac{1}{2}\mathcal{R} g_{ab}.
\end{equation}
In order to bring a generic perturbation to the Lorentz gauge we have to solve:
\begin{equation}
	 \xi_{a\,\,\, ;b}^{\,\,;b}
+\xi^b_{\, \, ;ba}+\frac{1}{2}\mathcal{R} \xi_a=h_{a\,\,;b}^{\,\,\,b},
\end{equation}
and in terms of the aforementioned scalar fields, 
\begin{eqnarray}
	\xi_a&=&\phi_{,a}+\epsilon_{ad}\psi^{,d} \quad \Rightarrow \quad \xi^b_{\,\,;b}=\Box\phi  \quad \Rightarrow \quad \xi^b_{\,\, ;b a} =(\Box \phi)_{,a}\\
	 \xi_{a\,\,\, ;b}^{\,\,;b}&=& g^{bc}\xi_{a;bc}= g^{bc} (\phi_{;abc}+\epsilon_a^{\,\, d}\psi_{;dbc})=g^{bc}(\phi_{;bac}+\epsilon_a^d \psi_{;bdc})
\end{eqnarray}

Recalling that for any vector $\lambda_a$
\begin{equation}
	\lambda_{b;ac}-\lambda_{b;ca}=\mathcal{R}^d_{bac}\lambda_d,
\end{equation}
then
\begin{equation}
	 \xi_{a\,\,\, ;b}^{\,\,;b}=(\Box \phi)_{,a} +\mathcal{R}_a^d \phi_{,d}+\epsilon_a^d(\Box \psi)_{,d}+\epsilon_a^d\mathcal{R}^l_d\psi_{,l}.
\end{equation}

In view of \eqref{eq:Rfirst} and the fact that the curvature is constant,
\begin{equation}
	\left[2\Box \phi+\frac{\mathcal{R}}{2}\phi\right]_{,a} +\epsilon_a^d\left[(\Box \psi)+\frac{\mathcal{R}}{2} \psi \right]_{,d}=v_a,
\end{equation}
where $v_a=h_{a\, ;b}^{\,\,;b} $.  Denoting,
$	\Phi=2\Box \phi +\frac{\mathcal{R}}{2}\phi,\,
	\Psi=\Box \psi+\frac{\mathcal{R}}{2} \psi$
results in
\begin{equation}
	\Phi_{,a} +\epsilon_a^b \Psi_{,b} =v_a.
\end{equation}

Since this is a general decomposition of a vector $v_a$ in two dimensions, we can always solve for  $\Phi,\Psi $ and take these two functions as source terms in their above definitions. Making a long story short, it is always possible to implement the Lorentz gauge, and the wave equation simplifies to
\be
\fn{\mathcal{R}^{(1)}}{\gamma_{ab},h_{ab}} =\Box \tilde h+\frac{1}{2}\tilde h \mathcal{R}^{(0)}(\gamma_{ab})=0.
\ee
 
 Solving the perturbation and backreaction equations in the Lorentz gauge in a background metric we will be interested in $\gamma_{ab}=Diag\{-p(r),1/p(r)\}$, turns out to be quite cumbersome and we shall not pursue it further.

\subsection{Traceless Gauge}
The traceless gauge is the second gauge that naturally emerges from Equation \eqref{eq:R1general}. Can we implement it? The trace  transforms as

\begin{equation}
	\tilde h'=\tilde h-2\xi'^a_{;a}.
\end{equation}

Starting from a general perturbation, one can reach the traceless gauge by solving  (see Equation \eqref{eq:xiv})
\begin{equation}
	\Box \phi=\frac{\tilde h}{2}
\end{equation}
for any $\psi$ in terms of the two scalar decomposition fields. The wave equation simplifies to
\be
\fn{\mathcal{R}^{(1)}}{\gamma_{ab},h_{ab}} =h^{ab}_{\quad;ab}=0,
\ee
and $h_{ab}$ is traceless. Can we implement the Lorentz and traceless gauges simultaneously? 
Suppose we first  bring the perturbation to the traceless gauge and then try to implement a Lorentz gauge. The former should not be disrupted by any further transformation.  Accordingly from the previous discussion, the traceless condition constrains  $\Box \phi=0 $. Then, implementing the Lorentz condition requires 
\begin{equation}
	\frac{\mathcal{R}}{2}\phi_{,a} +\epsilon_a^b \Psi_{,b} =v_a,
\end{equation}
with  $\Psi$ as before but $\phi$ satisfies 
\begin{equation}
	  \Box \phi=0.
\end{equation}

Clearly these two conditions cannot be met simultaneously.  Fulfillment of both gauge conditions would mean that all degrees of freedom can be gauged away. The traceless condition by itself leaves two degrees of freedom. 
\footnote{For example, schematically, the perturbed part can still be written as 
\be
h_{ab} = 
\begin{pmatrix} 
\fn{h}{t,r} & h_{01}(t,r) \\ 
h_{01}(t,r) & \fn{h}{t,r} 
\end{pmatrix}, 
\nonumber \end{equation} that has two independent entries.}
We can still remove another one. 
We require $h'_{01}=0$, while keeping the perturbation traceless, i.e., $\square \phi=0$ again,
\begin{eqnarray}
	h'_{01}=h_{01}-\xi_{0,1}-\xi_{1,0}+2\Gamma^a_{01}\xi_a=0.
\end{eqnarray}

Using \eqref{eq:xiv}, the fact that $\Gamma^b_{cb}=\partial_c\log \sqrt{-g}$ and some manipulations yields $h'_{01}=0$, and a traceless perturbation provided that:
\bea
\square \phi&=&0,\cr
\square \psi&=&\frac{h_{01}-2\phi_{;01}}{\sqrt{-g}}.
\eea

Hence, we can always solve the equation for $\phi$ and use it together with $h_{01}$ as a source for the equation for $\psi$. To summarize, a general perturbation can always be brought to a traceless diagonal form. In Section \ref{Marcelo1} we shall analyze the waves and backreaction generated in the traceless gauge.

\subsection{Traceful Gauge}
As our next section shows, trapping is very natural and intuitive in a special traceful gauge. The full curved spacetime analysis is rather cumbersome, so we limit our gauge discussion here to the flat spacetime case. We expect the gauge to be well-posed and completely fixed also in curved spacetime.
Consider the Minkowski background metric.
Starting from a generic perturbation, we would like to work in a gauge where the perturbed metric can be written as $g_{ab}=Diag\{-1+h,1-h\}$. In such a case, the gauge transformations we need to solve are:
\bea
h=h_{00}-2\xi_{0,0}\label{eq:1flat}\\
-h=h_{11}-2\xi_{1,1}\label{eq:2flat}\\
0=h_{01}-\xi_{0,1}-\xi_{1,0}\label{eq:3flat}
\eea

From the last equation we have
\be
\xi_1=\int dt\left(h_{01}-\xi_{0,1}\right)+f_1(x)
\ee

By construction $h=-h_{11}+2\xi_{1,1}$. Substituting into \eqref{eq:1flat} gives:
\be
h_{00}-2\xi_{0,0}=-h_{11}+2\xi_{1,1}=-h_{11}+2\left[\int dt\left(h_{01}-\xi_{0,1}\right)+f_1(x)\right]_{,1}
\ee

This is an inhomogeneous partial integro-differential equation. Let us differentiate the equation w.r.t. time:
\be
h_{00,0}-2\xi_{0,00}=-h_{11,0}+2(h_{01,1}-\xi_{0,11}) \quad \Rightarrow  \quad \square \xi_0=-\frac{h_{00,0}+h_{11,0}-2h_{01,1}}{2}
\ee

This is again the standard wave equation in 2D with a source term, that always has a solution. Since we have used up all the gauge freedom, it is a physical choice and not a gauge artefact. The perturbed part of the metric is  $h_{ab}=Diag\{h,-h\}$. We shall use this gauge in the upcoming section.
 
\section{Gravitational Waves Trapped in Space---Traceful Gauge}\label{GW}

\subsection{Background Geometry of the Vacuum Solution}

Consider the unperturbed metric 
\begin{equation}
\gamma_{ab} = 
\begin{pmatrix} 
-\fn{p}{r} & 0 \\ 
0 & \frac{1}{\fn{p}{r}}  
\end{pmatrix}. 
\label{eq:metric}
\end{equation}

 The equation of motion, Equation (\ref{eq:Einstein}), is then 
 \begin{equation}
\mathcal{R}^{(0)} = -\fn{p''}{r} = \Lambda
\label{eq:background}
\end{equation}
and the solution of  Equation (\ref{eq:Rzeroth}) is given by 
\begin{equation}
\label{eq:vacp}
\fn{p}{r} = A + Br - \frac{ \Lambda}{2} r^2,
\end{equation}
where $A$ and $B$ are constants.\footnote{If $\Lambda=0$ then the Riemann tensor vanishes, and one can rewrite the metric in the standard Minkowski form.}
 This is similar to the dS solution in static coordinates in 4D if we suppress the angular part. 

\subsection{Trace Wave Perturbations}

Let us now consider the perturbed metric 
\begin{equation} \label{eq:metricperturbed}
g_{\mu \nu} = 
\begin{pmatrix} 
-\fn{p}{r}+\fn{h}{t,r} & 0 \\ 
0 & \frac{1}{\fn{p}{r}} - \fn{h}{t,r} 
\end{pmatrix} 
\end{equation}
as in \cite{Sikkema:1989ib}.

A geon would have the form of 
\begin{equation}
\fn{h}{t,r} = \fn{T}{t} \fn{R}{r},
\end{equation}
where the time part would be $\fn{T}{t} \propto e^{-i\omega t}$ and the spatial part $\fn{R}{r}$ should be confined in space. 
Equation (\ref{eq:Rfirst}), then reads: 
\begin{equation}
R'' - \left( \frac{p'}{2p} + \frac{pp'}{2}  \right) R' 
- \left( \frac{p'^2}{2}+pp''+\frac{p''}{p} -\frac{p'^2}{2p^2} -\omega^2 \right) R = 0,
\label{eq:R}
\end{equation}
where prime denotes a derivative with respect to $r$.
We expect that the possibility of trapped waves would be checked by exploring the form of asymptotic behavior of \mbox{Equation (\ref{eq:R})} with a given $\fn{p}{r}$. We find two asymptotic behaviors as follows.

[{\bf AB1}]: 
 The first asymptotic behavior is that  {\it the waves can be trapped 
 in the region where $p \rightarrow 0$.}
For $p \rightarrow 0$, Equation (\ref{eq:R}) becomes
\begin{equation}
R'' - \frac{p'}{2p} R'- \left( \frac{p''}{p} -\frac{p'^2}{2p^2} \right) R = 0
\label{eq:Rpzero}
\end{equation} 

We may put
\begin{equation}
\fn{p}{r} = -\frac{\Lambda}{2} \left( r - \alpha \right) \left( r - \beta \right)
\end{equation}
and Equation (\ref{eq:Rpzero}) bcomes
\begin{equation}
R'' - \frac{1}{2}\left( \frac{1}{r-\alpha} + \frac{1}{r-\beta}  \right) R' 
+ \left[\frac{1}{2 \left( r-\alpha \right)^2} + \frac{1}{2 \left( r-\beta \right)^2} -\frac{1}{\left( r-\alpha \right)\left( r-\beta \right)}  \right] R = 0.
\end{equation}

Without loss of generality, we can consider the asymptotic behavior of the solutions around $r= \alpha$. The solution is 
\begin{equation}
\fn{R}{r} = \left(r- \alpha \right)^{\frac{1}{2}} \left\{ C_1  \exp \left[ 2\left( \frac{r-\alpha}{\alpha-\beta} \right)^\frac{1}{2} \right]
+ C_2 \exp \left[ -2\left( \frac{r-\alpha}{\alpha-\beta} \right)^\frac{1}{2} \right] \right\}
\end{equation}
which means the solution $R(r)$ becomes zero as $p \rightarrow 0$ for $r \rightarrow \alpha\,$. \footnote{An interesting situation occurs if there is a single root, i.e. 
$\alpha=\beta$ in region {\bf AB1}. In such case, the lowest order approximation becomes a Bessel-type equation: $R''- \frac{p'}{2p} R'+\omega^2 R=0$, with the solution $(r-\alpha)\left[c_1J_1(-\omega(r-\alpha))-c_2Y_1(-\omega(r-\alpha))\right]$. In such a case, the limit $r\rightarrow \alpha$ can actually be finite with $\lim_{r\rightarrow \alpha}R=\frac{2c_2}{\pi \omega}$.}

[{\bf AB2}]: The second asymptotic behavior is that {\it the waves cannot be trapped 
 in the region where $p \rightarrow \pm \infty$}. 
For $p \rightarrow \pm \infty$, 
Equation (\ref{eq:R}) becomes
\begin{equation}
R'' - \frac{pp'}{2} R' 
- \left( \frac{p'^2}{2}+pp''\right) R = 0 \label{eq:Rpinfty}
\end{equation}

In this case, when $\fn{p}{r} = A + Br - \frac{ \Lambda}{2} r^2$, 
the solution is given by
\begin{equation}
\label{eq:Solinfty}
\fn{R}{r} =  C_1 \fn{p'}{r} \fn{F}{r} + C_2 \fn{p'}{r} \fn{F}{r} \int \frac{ 1}{\fn{p'}{x}^2 \fn{F}{x} } dx  ,
\end{equation}
where 
\begin{equation}
\fn{F}{r} = \exp\left[ \frac{\fn{p}{r}^2}{4}  \right]
\end{equation}

In Equation (\ref{eq:Solinfty}), $C_1$ term is definitely divergent. The  $C_2$ term diverges or goes to zero as $r \rightarrow \infty$ on very particular cases such as a single degenerate root. Nevertheless, waves extending from some root of $p(r)$ to infinity cannot be considered as finite and localized.

These two conditions seem simple but predict where the waves can be confined in space. If $R(r)$ has some finite support, then we get trapping. If not, then we cannot say that the waves are confined.
We will turn back to this point in \sref{numerical}. 

\subsection{Backreaction Analysis}

The backreaction of waves on the vacuum metric is calculated by  Equation (\ref{eq:Rsecond}) which reduces to 
\begin{eqnarray}
\begin{aligned}
2\tilde{p}^3 \left(\tilde{p}''+\Lambda \right)   = \left \langle  2h^2\tilde{p}'^2-2h^2\tilde{p}^4\tilde{p}'^2 -2h^2\tilde{p}\tilde{p}''-2h^2\tilde{p}^3\tilde{p}''  
-2h^2\tilde{p}^5\tilde{p}''-\tilde{p}^2\dot h^2  -\tilde{p}^4\dot h^2 \right. \\ 
\left. -2h\tilde{p}^2\ddot h  -2h\tilde{p}^4\ddot h-3h\tilde{p}\tilde{p}'h' -h\tilde{p}^3\tilde{p}'h'  -2h\tilde{p}^5\tilde{p}'h'+\tilde{p}^2h'^2 +\tilde{p}^4h'^2+2h\tilde{p}^2h''+2h\tilde{p}^4h'' \right \rangle 
\end{aligned}\label{eq:Rsecond_h}
\end{eqnarray} 
where $\fn{p}{r}$ is modified into $\fn{\tilde{p}}{r}$ and a dot dentoes a derivative with respect to time.  
After considering $\langle (e^{i \omega t})^2 \rangle = \frac{1}{2} $,  Eqaution (\ref{eq:Rsecond_h}) becomes
\begin{eqnarray}
\begin{aligned}
2\tilde{p}^3\left(\tilde{p}''+\Lambda \right) =   \left(  \tilde{p}'^2   -  \tilde{p}^4\tilde{p}'^2  - \tilde{p} \tilde{p}''  - \tilde{p}^3\tilde{p}''    
-\tilde{p}^5\tilde{p}'' + \frac{3\omega^2}{2} \tilde{p}^2   + \frac{3\omega^2}{2} \tilde{p}^4      \right) R^2   \\ 
- \left( \frac{3}{2} \tilde{p}\tilde{p}' + \frac{1}{2} \tilde{p}^3\tilde{p}' + \tilde{p}^5\tilde{p}' \right) RR' 
+ \left( \frac{1}{2} \tilde{p}^2  + \frac{1}{2} \tilde{p}^4 \right) R'^2 + \left( \tilde{p}^2 + \tilde{p}^4 \right) RR'' ~.
\end{aligned}\label{eq:Rsecond_final}
\end{eqnarray} 

It is difficult to find analytic solutions for Equation (\ref{eq:Rsecond_final}) and we will try to solve it by numerical simulations. 
However, we can mention two main features that would be reflected in the numerical results:
First, when $R \ll 1$ and $R' \ll 1$, Equation (\ref{eq:Rsecond_final}) gives $\tilde{p}''+\Lambda \simeq 0$ reproducing Equation (\ref{eq:background}). Hence, the background geometry will not change and $p$ is nearly the same as $\tilde{p}$.
Second, when $\omega \gg 1$, $\omega$ terms get important in the right hand side of Equation (\ref{eq:Rsecond_final}). One may expect that the mode of large $\omega$ cause substantial backreaction to the background metric.

\subsection{Numerical Results} \label{numerical}

\begin{figure}[H]
\includegraphics[width=0.9\linewidth]{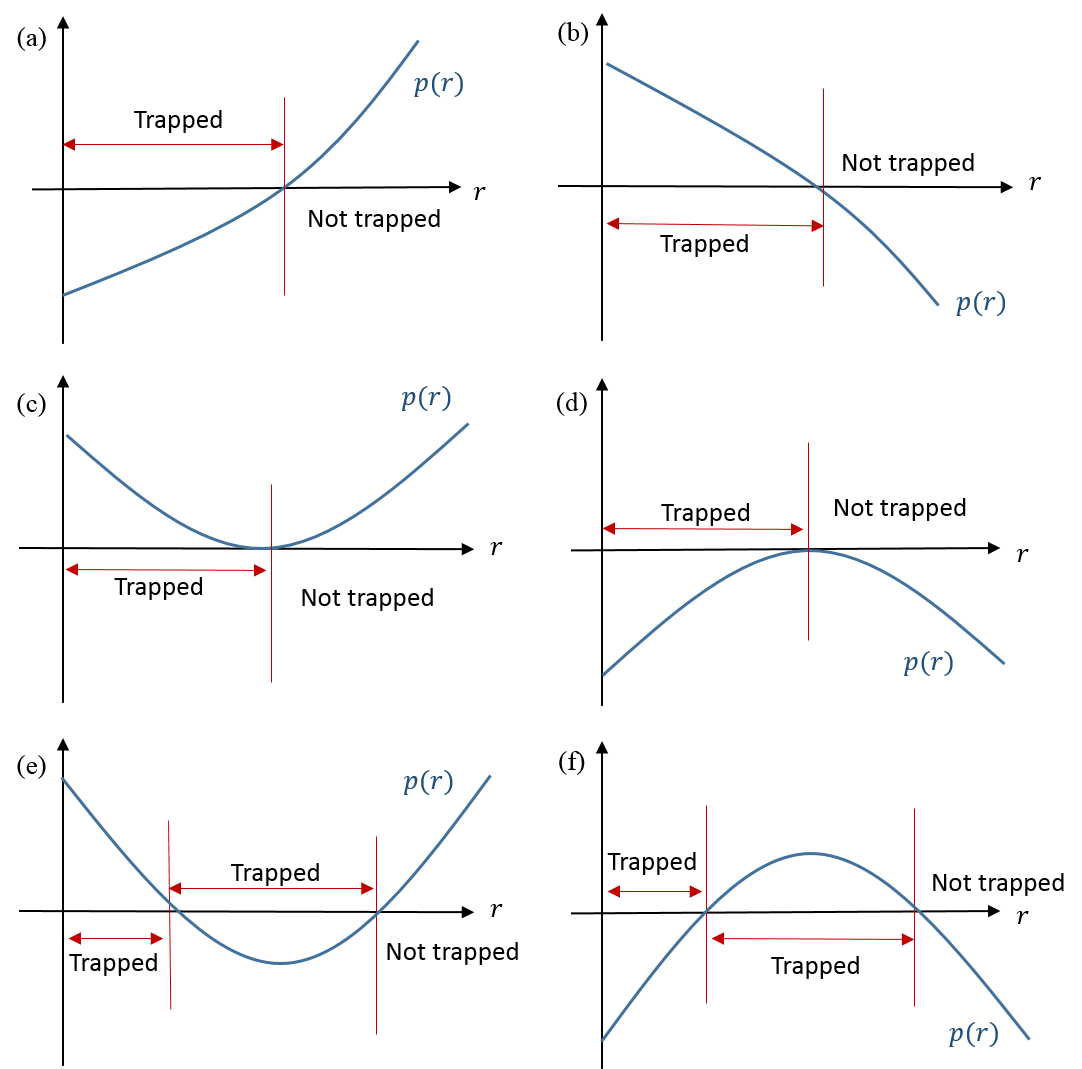}
\caption{Schematic 
 plots showing GWs are trapped with respect to the shape of p(r). The plot does not change qualitatively even after including the backreaction resulting in $\tilde p(r)$.}
\label{fig:schematic}
\end{figure}

In JT gravity, the metric component $\fn{p}{r}$ is presented by quadratic curves, as given in Equation (\ref{eq:vacp}).  The global structure of spacetime represented by the signs of $\Lambda$ is not crucial to decide whether there are trapping regions.
Rather, the existence and positions of zeros of $\fn{p}{r}$ are critical in predicting the trapping regions from the asymptotic behaviors of the analytic solutions of Equation (\ref{eq:R}), AB1 and AB2. Note that the physically proper range should be $r > 0$ from our metric ansatz. 
The waves would be trapped between points where the metric component $p(r)$  goes to zero but not in regions where it becomes divergent. Borrowing from a 4D language, they are trapped between ``horizons''---either inside the inner horizon, or between the inner and outer horizons. 
This condition applies to all the cases regardless of the values of $\Lambda$. It looks straightforward but is useful to clarify the cases.  
It is valid for $\fn{\tilde{p}}{r}$ even after considering the backreaction to the background through Equation (\ref{eq:Rsecond_final}). This general statement is schematically summarized in Figure \ref{fig:schematic}. 

Since $p(r)$ is dimensionless, we have two dimensionful parameters $B$ and $\Lambda$. Except for the $B=0$ case, we present our results in units where $B=1$ due to numerical limitations. As an order of magnitude estimate, in the absence of $\Lambda$ the horizon is at 
 $r=-B/A$. Both $B^{-1}$ and $\Lambda^{-1/2}$ have arbitrary units of length. Hence, all our results are presented in such arbitrary units of length.

\begin{figure}[H]
\includegraphics[width=0.8\linewidth]{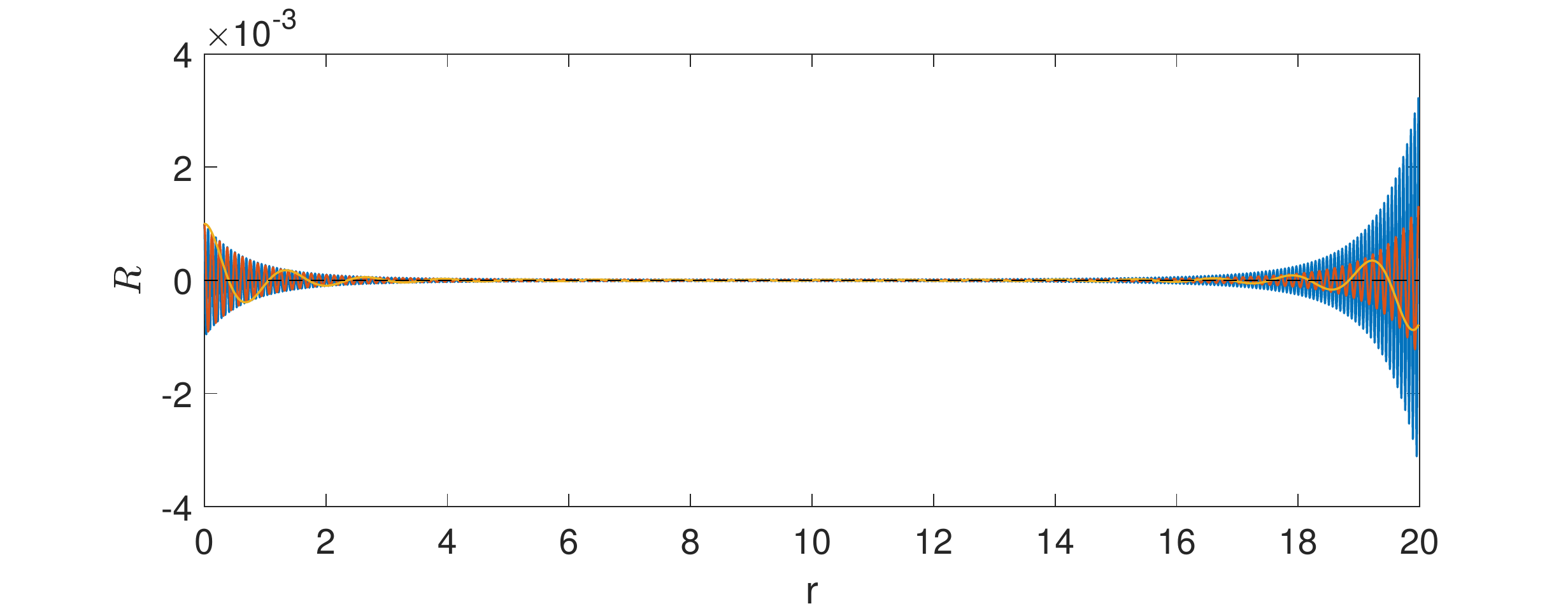}
\includegraphics[width=0.8\linewidth]{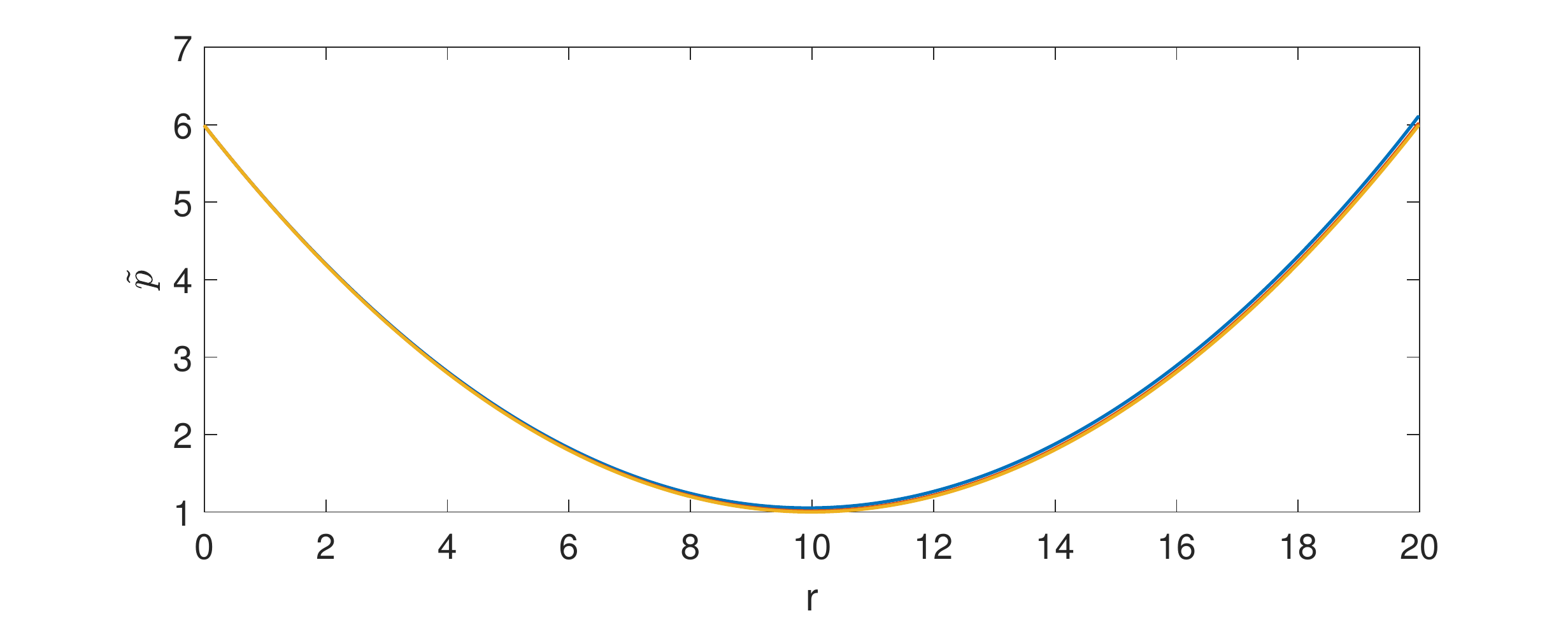}
\caption{Top: 
 $\fn{R}{r}$ with $\omega = {5},  {50}, {100}$ (orange, red and cyan curves respectively) for $\fn{p}{r}=\frac{\left(r-10\right)^2}{20}+1 , R(0.01)=0.001, \fn{R'}{0}=0$.
Bottom: $\fn{\tilde{p}}{r}$ with $\omega = {5},  {50}, {100}$ (orange, red and cyan curves respectively, i.e. bottom to top) considering the backreaction to $\fn{p}{r}=\frac{\left(r-10\right)^2}{20}+1$. $\tilde p(r)$ is strictly positive, so no GWs are trapped anywhere. {$B=-1$} units are used.}
\label{fig:Lm0.2}
\end{figure}

Let us discuss the different cases. First, if there exists no zero of $\fn{p}{r}$ for $r \geq 0$, waves cannot be trapped.
We describe one of these cases in Figure \ref{fig:Lm0.2} where $\fn{p}{r}=\frac{1}{20}\left(r-10\right)^2+1$ is considered. Waves are all divergent as $r \rightarrow \infty$ as can be seen in the top-panel of Figure \ref{fig:Lm0.2}. 
 In the bottom, we present as $\tilde p(r)$, i.e., the modification of the background metric due to the backreaction. The backreaction increases with $\omega$, and it gets more challenging to confine waves in space. 
 
\begin{figure}[H]
\includegraphics[width=0.9\linewidth]{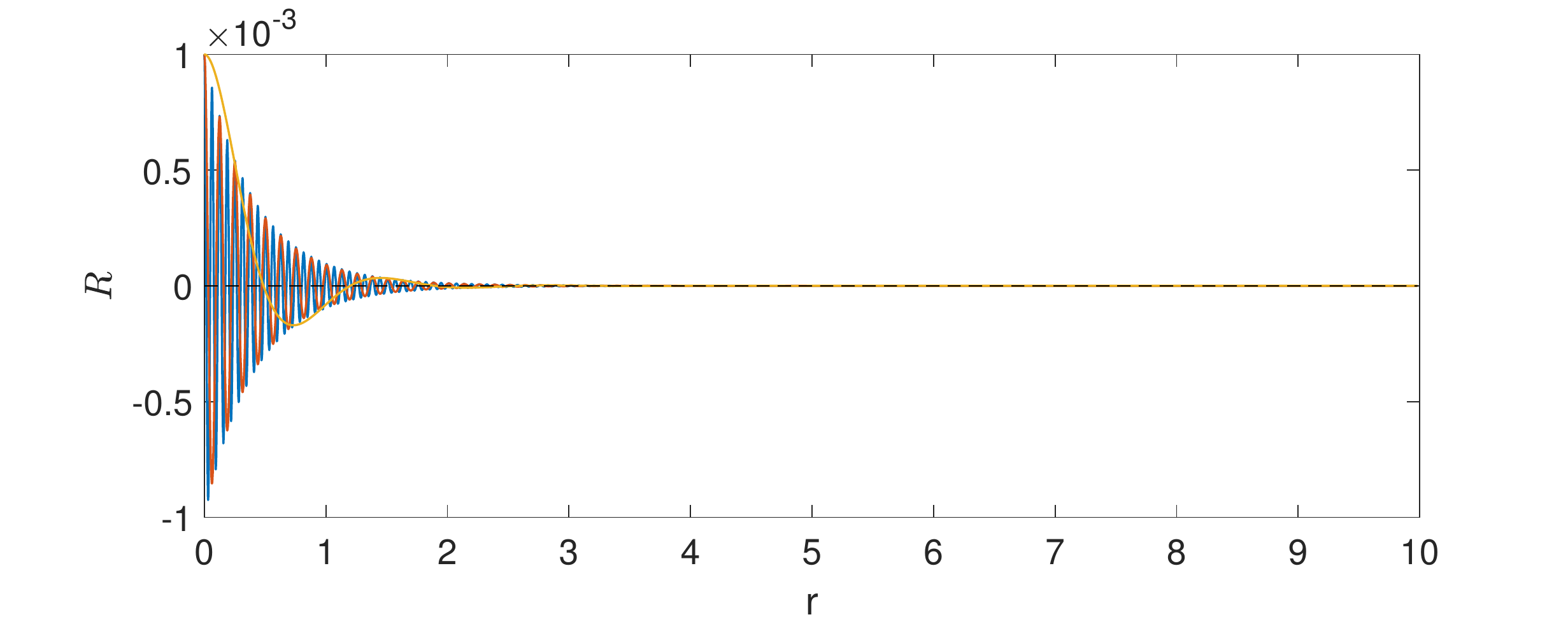}
\includegraphics[width=0.9\linewidth]{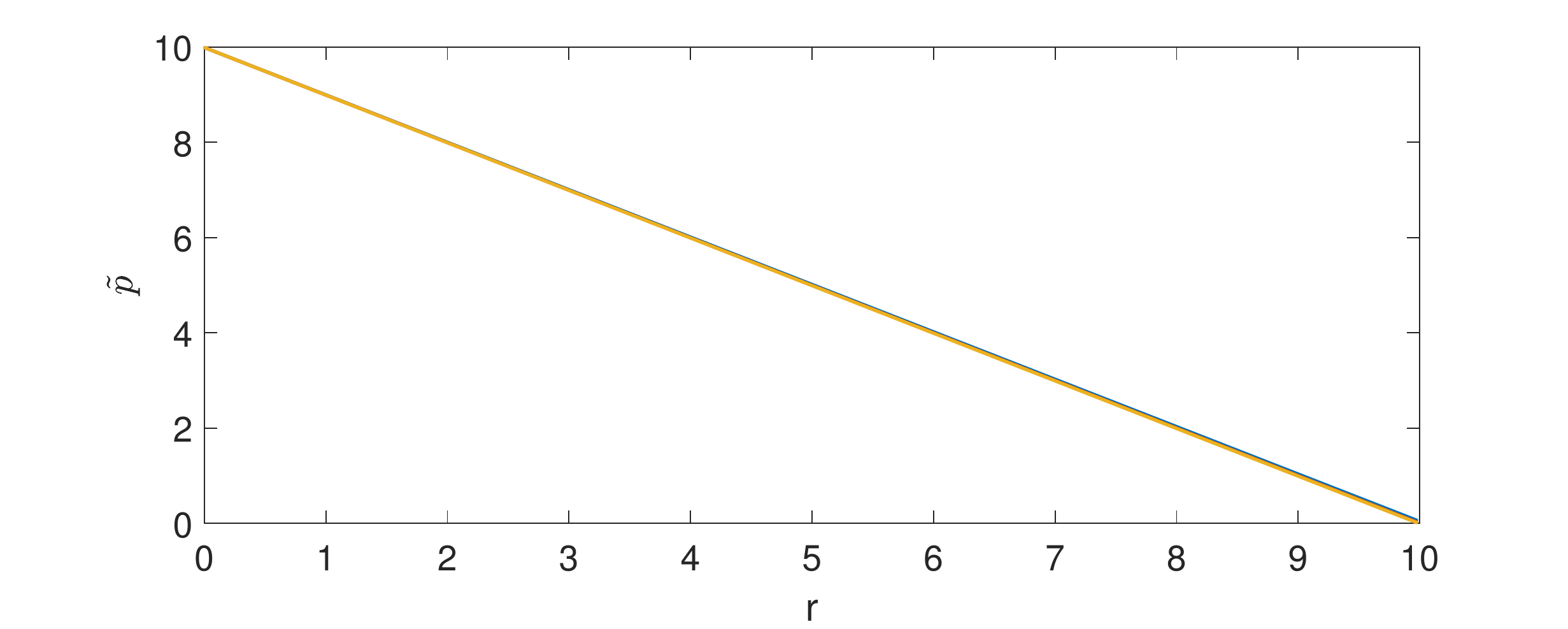}
\caption{Top: $\fn{R}{r}$ with $\omega = {5},  {50}, { 100}$  (orange, red and cyan curves respectively) for $\fn{p}{r}=10-r , R(0.01)=0.01, \fn{R'}{0}=0$.
Bottom: $\fn{\tilde{p}}{r}$ with $\omega = {5},  {50}, { 100}$ (orange, red and cyan curves respectively, i.e. bottom to top) considering the backreaction to $\fn{p}{r}=10-r$. {$B=-1$} units are used.}
\label{fig:L=0}
\end{figure}

\begin{figure}[H]
\includegraphics[width=0.8\linewidth]{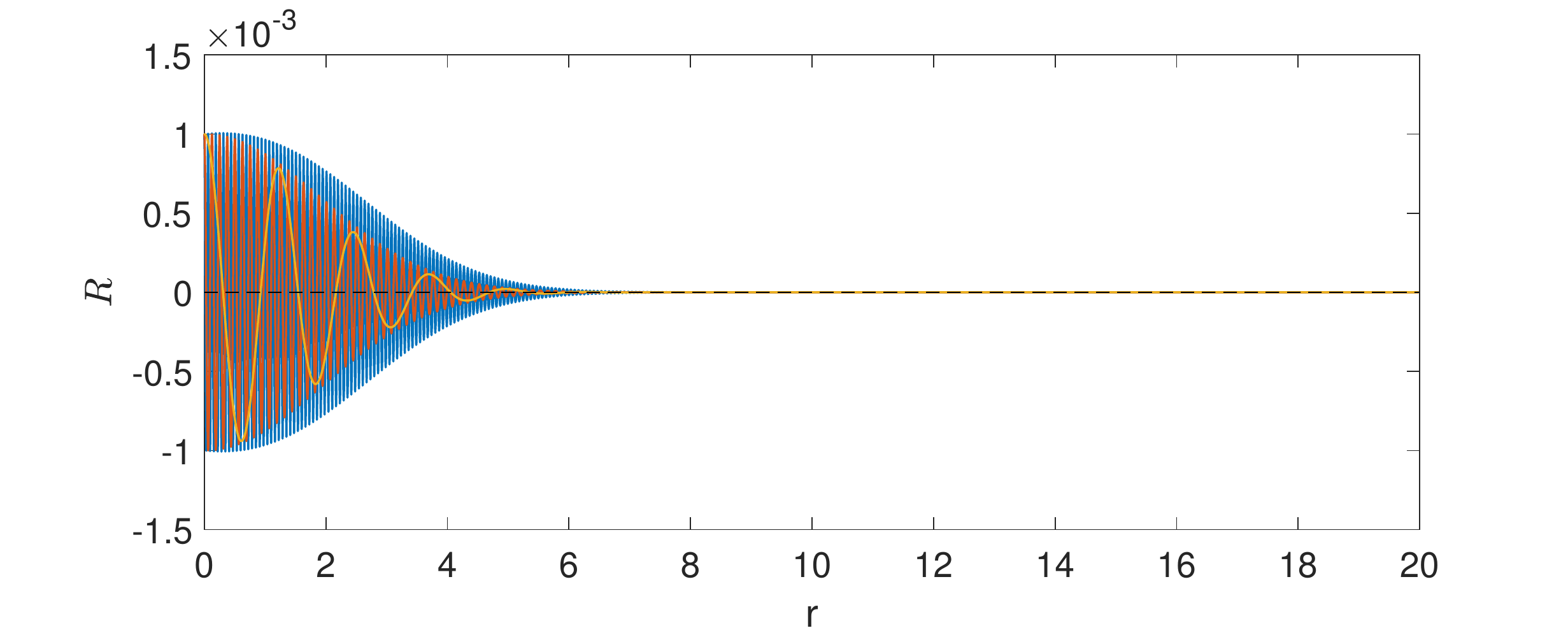}
\includegraphics[width=0.8\linewidth]{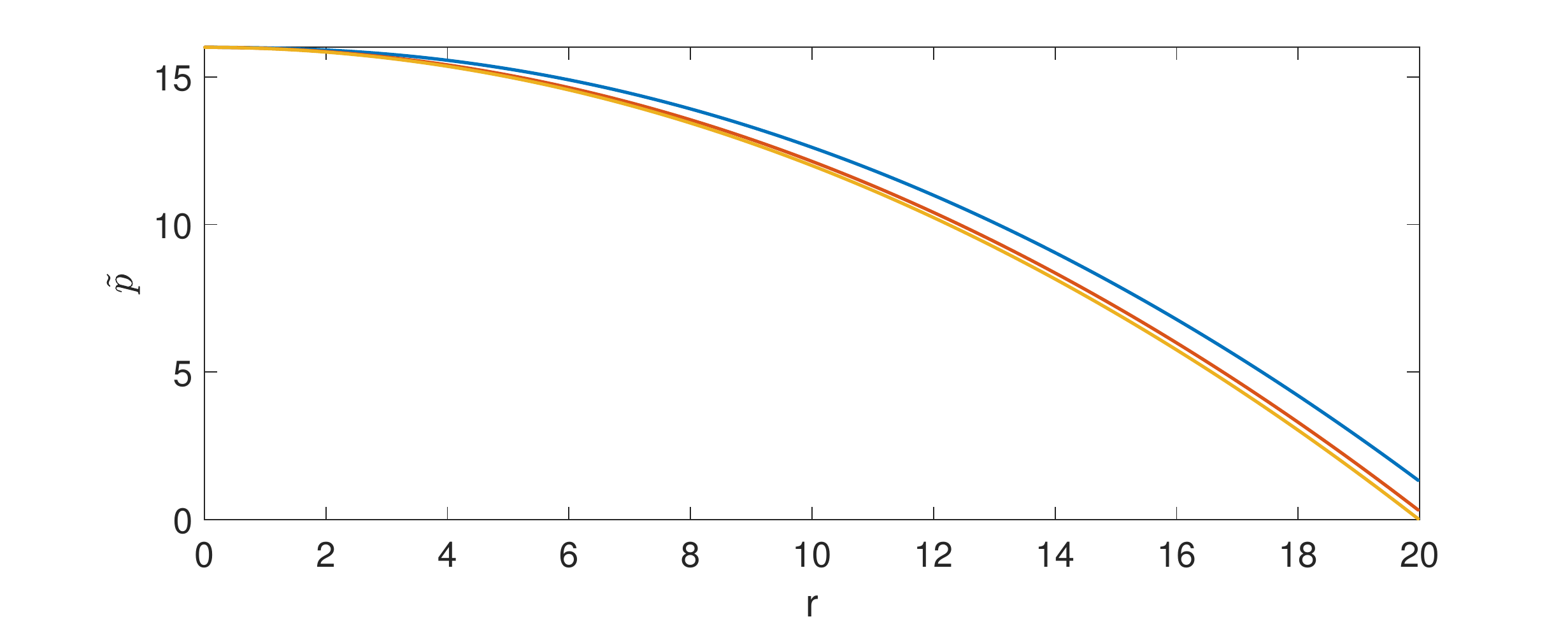}
\caption{Top: $\fn{R}{r}$ with $\omega = {5},  {50}, { 100}$ (orange, red and cyan curves respectively) for $\fn{p}{r}=-\frac{r^2}{25}+16 , R(0.01)=0.01, \fn{R'}{0}=0$.
Bottom: $\fn{\tilde{p}}{r}$ with $\omega = {5},  {50}, {100}$ (orange, red and cyan curves respectively, i.e. bottom to top) considering the backreaction to $\fn{p}{r}=-\frac{r^2}{25}+16$. Arbitrary units of length such that $\Lambda=0.08 \, a.u.^{-1/2}$.}
\label{fig:L0.2}
\end{figure}

 Second, if there exists one zero $r_1$ of $\fn{p}{r}$ for $r \geq 0$, waves can be trapped in the region where $\fn{p}{r}$ is finite or does not go to the infinity (i.e., $0<r<r_1$). 
Two examples are given: $\fn{p}{r}=10-r$ with $\Lambda = 0$ in Figure \ref{fig:L=0} that mimics a black hole horizon, and $\fn{p}{r}=-\frac{r^2}{25}+16$ with $\Lambda < 0$ in Figure \ref{fig:L0.2}, that mimics a cosmological dS horizon. 

\begin{figure}[H] 
\includegraphics[width=0.9\linewidth]{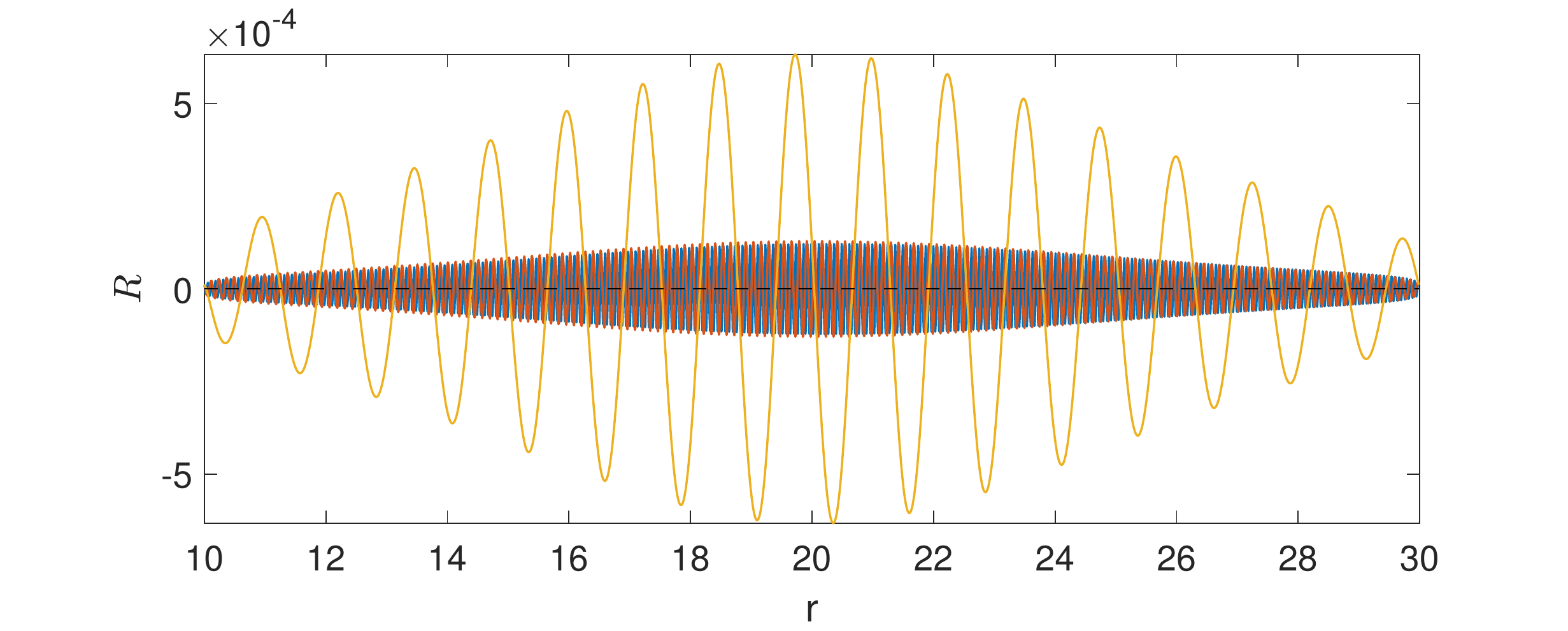}
\includegraphics[width=0.9\linewidth]{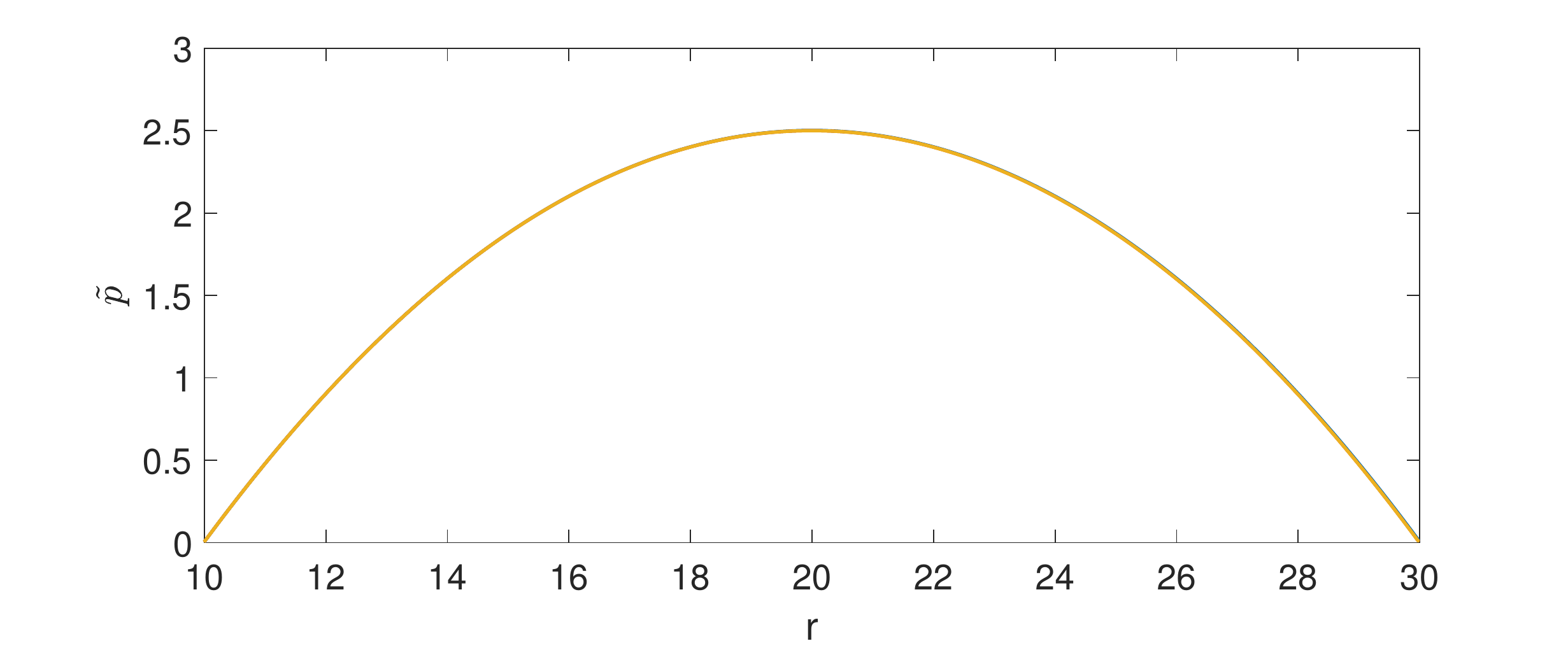}
\caption{Top: $\fn{R}{r}$ with $\omega = {5},  {50}, {100}$ (orange, red and cyan curves respectively) for $\fn{p}{r}=-\frac{1}{40}\left(r-10\right)\left(r-30\right) , R(10.01)=0.00001, \fn{R'}{0}=0$. {The GW are trapped between the ``Schwarzschild horizon'' and the ``dS horizon''.}
Bottom: $\fn{\tilde{p}}{r}$ with $\omega = {5},  {50}, {100}$ (orange, red and cyan curves respectively, i.e. bottom to top) considering the backreaction to $\fn{p}{r}=-\frac{1}{40}\left(r-10\right)\left(r-30\right)$. $B=1$ units are used.}
\label{fig:standing_desitter}
\end{figure}

Third, if there exist two zeros of $\fn{p}{r}$ for $r \geq 0$, waves can be trapped in two regions, between the origin and the smaller zero ($0 < r< r_1$), and between the zeros ($r_1 < r < r_2$). 
The reasoning of Figures \ref{fig:L=0} and \ref{fig:L0.2}  applies to the trapping region $0<r<r_1$, and hence we focus on $r_1 < r < r_2$ which seems more interesting.  
In Figure \ref{fig:standing_desitter}, we present such an example with $\fn{p}{r}=-\frac{1}{40}\left(r-10\right)\left(r-30\right)$ whose two zeros are $r_1=10,\, r_2=30$ and the CC is positive $\Lambda > 0$. Waves are confined in the region of $10 < r < 30$, { i.e. they are trapped between the ``Schwarzschild horizon'' and the ``dS horizon''.} Notice that the backreaction to the vacuum metric in the case of two zeros of $p(r_1)=p(r_2)=0$, is not as substantial and $\fn{\tilde{p}}{r}$ does not change severely compared to the case of Figure \ref{fig:Lm0.2} or even the single zero case of Figures \ref{fig:L=0} and  \ref{fig:L0.2}. 
What happens if $\fn{p}{r} = 0$ has a single degenerate root? Waves are trapped in the region between the origin and the root. It is essentially the same as the cases of Figures \ref{fig:L=0} and  \ref{fig:L0.2}.

\vspace{-6pt}
\vspace{-6pt}

\section{Traceless Gravitational Waves Perturbations}\label{Marcelo1}
 Let us now consider the following GW perturbations, where we are manifestly in the weak field limit $h(t,r)\ll1$ as well as $h^{a}_{\,a}=0$. We raise and lower indices with the background metric. We shall see that the equations simplify considerably and are amenable to analytic solutions as well. The metric now reads:
\begin{equation}
g_{ab} = 
\begin{pmatrix} 
-\fn{p}{r}(1-\fn{h}{t,r}) & 0 \\ 
0 & \frac{1}{\fn{p}{r}}(1 + \fn{h}{t,r}) 
\end{pmatrix} .
\end{equation}

Considering again,
$\fn{h}{t,r} = \fn{T}{t} \fn{R}{r}$,
with periodic time dependence, $\fn{T}{t} \propto e^{-i\omega t}$, 
the first order GW equation is
\begin{equation}
R'' + 2\frac{p'}{p} R' 
+\left(\frac{p''}{p}-\frac{\omega^2}{p^2}\right) R = 0.
\end{equation}

We can rewrite the equation in a very simple form for u(r)=p(r) R(r) as
\be
u''-\frac{\omega^2}{p^2}u=0.
\ee

The general solution is given by
\be
R(r)=\sqrt{\frac{2}{p(r)}}\left(c_1e^{-s/d \arctan\left(\frac{B-\Lambda r}{d}\right)}+\frac{c_2}{2 s}e^{s/d \arctan\left(\frac{B-\Lambda r}{d}\right)}\right),
\ee
where $s=\sqrt{B^2+2A \Lambda+4\omega^2}, d=\sqrt{B^2+2 A \Lambda}$. Notice that we can absorb various constants into $c_1,c_2$ if we wish. To better understand the qualitative behavior, let us use the form $p(r)=-\frac{\Lambda}{2}(r-\alpha)(r-\beta)$, where $\beta>\alpha$ without loss of generality. Hence the solution is of the form
\be
R(r)=\frac{2}{\sqrt{p(r)}}\left(\tilde c_1 \left(\frac{r-\alpha}{r-\beta}\right)^{\frac{1}{2}\sqrt{1+\left[\frac{4\omega}{\Lambda(\alpha-\beta)}\right]^2}}+\tilde c_2 \left(\frac{r-\beta}{r-\alpha}\right)^{\frac{1}{2}\sqrt{1+\left[\frac{4\omega}{\Lambda(\alpha-\beta)}\right]^2}}\right).
\ee

The backreaction of gravitational waves to the vacuum metric is calculated by  \mbox{Equation (\ref{eq:Rsecond})} which reduces again to a very simple form
\begin{eqnarray}
\left(\tilde{p}''+\Lambda\right)=-  \left \langle 2 \tilde p'' h^2 +\tilde p' h h'\right \rangle
\eea

After averaging over time and using $\langle (e^{i \omega t})^2 \rangle = \frac{1}{2} $ we get:
\be
\left(1+\frac{R^2}{2}\right)\tilde p''+\frac{R R'}{2}\tilde p'=-\Lambda
\ee

Using the first order solution, this can actually be solved analytically:
\be
\tilde p(r)=c_1+\frac{c_2}{\sqrt{2}}\int^r\frac{dx}{\sqrt{1+R^2(x)/2}}-\frac{\Lambda}{2}\left[\int^r\frac{dx}{\sqrt{1+R^2(x)/2}}\right]^2
\ee

Two numerical examples are presented in Figures \ref{fig:marcelo1} and \ref{fig:marcelo2}. In Figure \ref{fig:marcelo1} we present trapping for the $\Lambda=0$ case, in units of $B=1$. In Figure \ref{fig:marcelo2} the trapping is in units of $\Lambda=1$ with the CC being negative unity. Unlike the trace waves, here the oscillatory behavior of the perturbation is only in time, and not in space. Worth noting is a significant backreaction on the metric as the frequency $\omega$ increases in the $\Lambda=0$ case, see the bottom panel of Figure \ref{fig:marcelo1}. 

\begin{figure}[H]
\begin{center}
\includegraphics[width=0.7\linewidth]{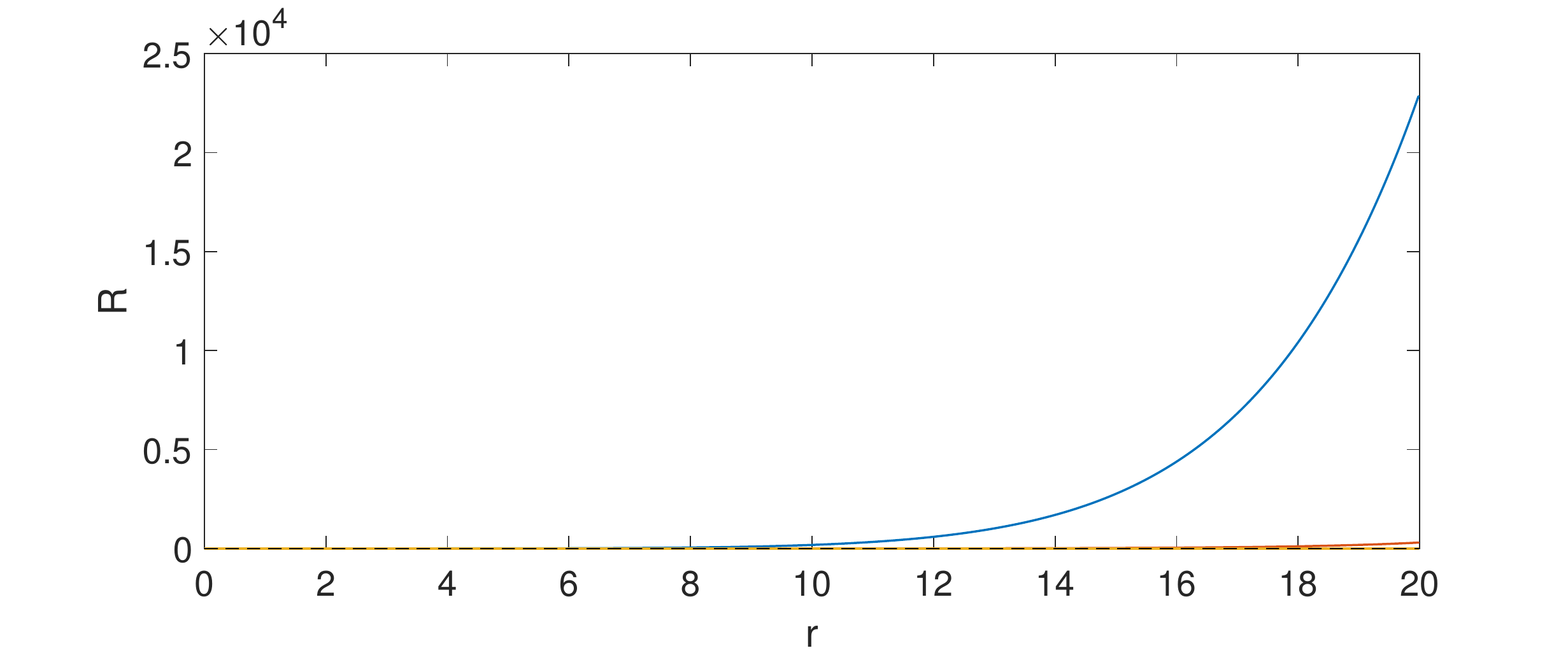}
\includegraphics[width=0.7\linewidth]{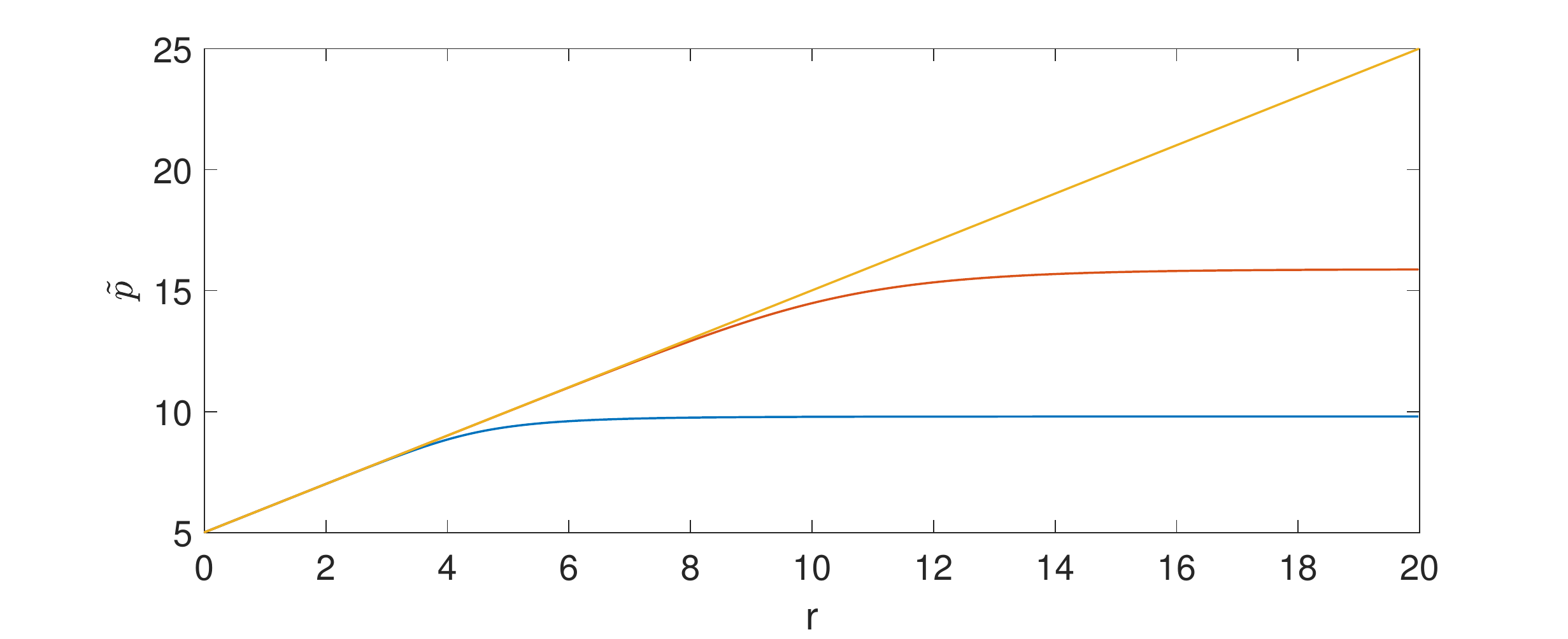}
\end{center}
\caption{Top: $\fn{R}{r}$ with $\omega ={1}, {5}, { 10}$  (orange, red and cyan curves respectively, i.e. bottom to top) for $\fn{p}{r}=5+r , R(0.01)=0.01, \fn{R'}{0}=0$.
Bottom: $\fn{\tilde{p}}{r}$ with $\omega ={1}, {5}, { 10}$ (orange, red and cyan curves respectively, i.e. top to bottom) considering the backreaction to $\fn{p}{r}=5+r$. Both plots are in the traceless gauge. $B=1$ units are used.}
\label{fig:marcelo1}
\end{figure}
\vspace{-6pt}
\begin{figure}[H]
\includegraphics[width=0.9\linewidth]{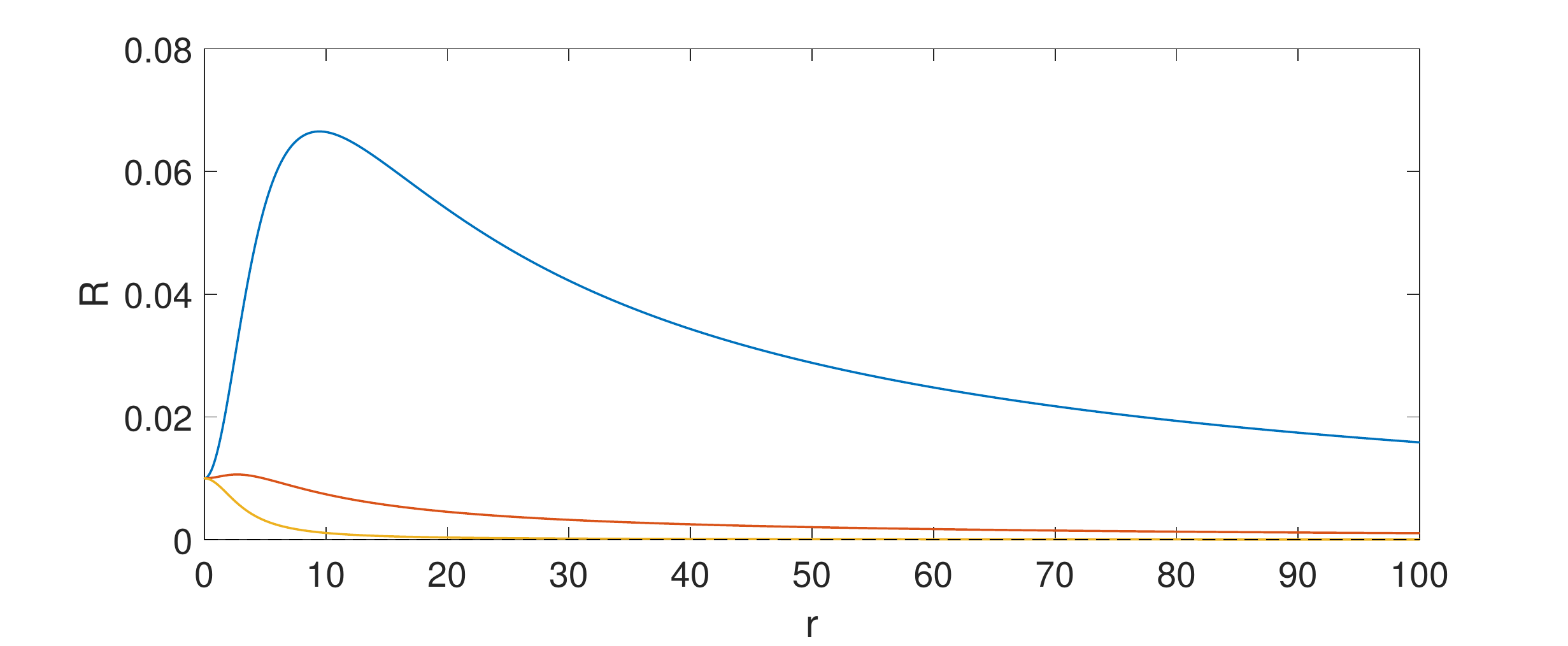}
\includegraphics[width=0.9\linewidth]{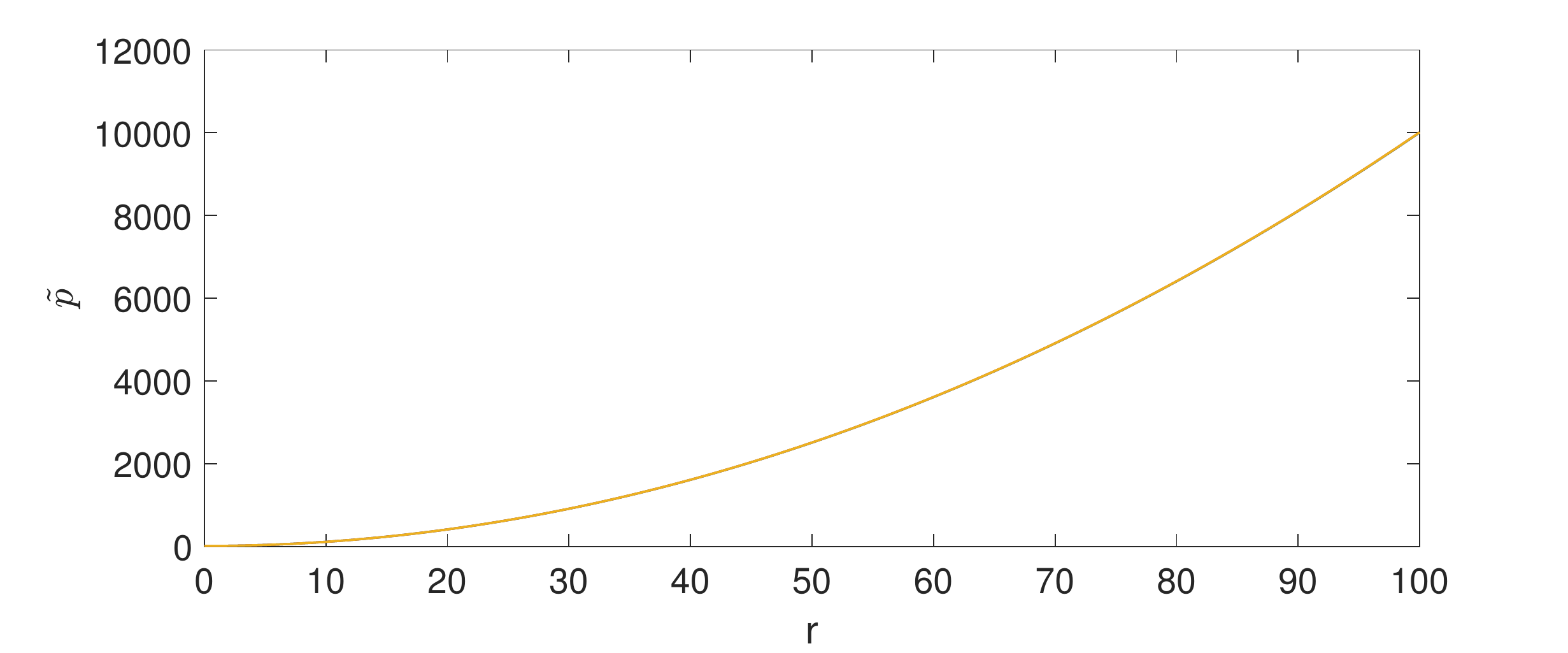}
\caption{Top: 
 $\fn{R}{r}$ with $\omega ={1}, {5}, { 10}$ (orange, red and cyan curves respectively, i.e. bottom to top) for $\fn{p}{r}=10+r^2 , R(0.01)=0.01, \fn{R'}{0}=0$.
Bottom: $\fn{\tilde{p}}{r}$ with $\omega ={1}, {5}, { 10}$ (orange, red and cyan curves respectively, i.e. bottom to top) considering the backreaction to $\fn{p}{r}=10+r^2$. Both plots are in the traceless gauge. $\Lambda=1$ units are used. }
\label{fig:marcelo2}
\end{figure}

\subsection*{Off-Diagonal Gauge}
Let us consider another traceless gauge which is off-diagonal, $h_{00}=h_{11}=0$, and $h_{01}\neq 0$. This is reminiscent of the $\times$ polarization in 4D. This gauge is obtained by direct integration of the first order PDEs that we have as a gauge transformation for our background ansatz.
Considering this gauge significantly simplifies the equation of motion which now reads:
\bea
\mathcal{R}^{(1)}=\frac{p'}{p}\partial_t h_{01}+2\partial_t \partial_r h_{01}=0 \Rightarrow
h_{01}=\frac{T(t)}{\sqrt{p(r)}}+f_2(r).
\eea

We can now consider the backreaction of the perturbation on the background metric:
\be
\mathcal{R}^{(2)}=\left(p'h_{01}\partial_r h_{01}+p''h_{01}^2\right).
\ee

Considering the $f_2(r)\equiv 0$ case, we can write down the backreaction equation, denoting the time average as $c\equiv \langle f(t)^2\rangle$:
\be
\mathcal{R}^{(0)}+\langle\mathcal{R}^{(2)}\rangle=-\tilde p''+c\left(\tilde p'h_{01}\partial_r h_{01}+\tilde p''h_{01}^2\right)=\Lambda,
\ee
with a well defined analytical solution:
\be
\tilde p(r)=c_1+c_2\int^r\frac{du}{\sqrt{1-c\, h_{01}(u)^2}}-\frac{\Lambda}{2}\left(\int^r\frac{du}{\sqrt{1-c\, h_{01}(u)^2}}\right)^2.
\ee

Now substituting the $h_{01}$ solution above gives:
\be
\int^r\frac{du}{\sqrt{1-c\, h_{01}(u)^2}}=\frac{\sqrt{B^2+2 A\Lambda}}{\Lambda}E\left(arcsin\frac{-B+\Lambda r}{\sqrt{B^2+2(A-c)\Lambda}}\,\Big{|}\,\frac{B^2+2(A-c)\Lambda}{B^2+2A\Lambda}\right),
\ee
where $E(\phi | m)$ is the elliptic integral of the second kind. So in this gauge the solution is fully analytic.

\section{Relation between Different Coordinate Systems} \label{synchronous}
Fortunately, the two-dimensional case is amenable to an exact analytical solution. Counting degrees of freedom, any spacetime in any dimensions can always be put into the synchronous form, and in the case of 2D into a conformally flat form. In particular, in such forms, we do not limit ourselves to a static ansatz plus a time-dependent perturbation. We first consider more general background solutions in three gauges: synchronous gauge, conformally flat gauge and spatially flat gauge. We then explicitly map our static ansatz to the latter two. Finally, we give an example of the perturbation and averaging analysis in the conformal gauge. 
Considering the synchronous gauge 
\begin{equation}
ds^2 =-d\tau^2 + F(r,\tau) dr^2,
\label{linelement}
\end{equation}
where $F(r,\tau)$ can be any function.
The solution of the JT vacuum equation of motion (EOM)
\begin{equation}
\mathcal{R} = \Lambda
\end{equation}
is given by
\begin{equation}
\fn{F}{r,\tau}=\fn{f}{r} \cosh\left[ \fn{g}{r} +\Lambda \tau  \right]
\end{equation}
where $\fn{f}{r}, \fn{g}{r}$ are arbitrary functions.

Similarly, considering the conformal gauge, 
\begin{equation}
ds^2 = \fn{G}{\eta,y}\left( -d\eta^2 + dy^2 \right).
\label{conformal}
\end{equation}
 
The EOM can be solved by 
\begin{equation}
\fn{G}{\eta,y}=-\frac{2}{\Lambda}\frac{c_1^2-c_2^2}{\fn{\cosh^2}{c_1 \eta+c_2 y+c_3}}
\end{equation}
where $c_1,c_2,c_3$ are determined by the boundary conditions. Let us stress, that this is a simple explicit solution, but does not describe all possible solutions.
Finally, considering a ``spatially flat'' gauge where
\be
ds^2 = - \fn{H}{\tilde t,r_*} d\tilde t^2 +dr_*^2
\ee
yields a solution that clearly exhibits an oscillatory propagating behavior:
\be
\fn{H}{\tilde t,r_*} = \fn{f_1}{\tilde t} \cos^2 \left[ \sqrt{\frac{\Lambda}{2}}\left( r_*-2 \fn{f_2}{\tilde t} \right)\right]
\ee
where again $\fn{f_1}{\tilde t}, \fn{f_2}{\tilde t}$ are arbitrary functions.  
Obtaining these exact solutions, let us demonstrate several numerical examples of spatial and time dependent periodical solutions in these gauges, i.e some form of ``waves'' of space-time. A propagating `soliton' in the conformal gauge is presented in  Figure \ref{fig:conformal}, and a wave packet that survives for a finite time in the spatially flat gauge in Figure \ref{fig:spatiallyflat}.

\begin{figure}[H]
\includegraphics[width=0.45\linewidth]{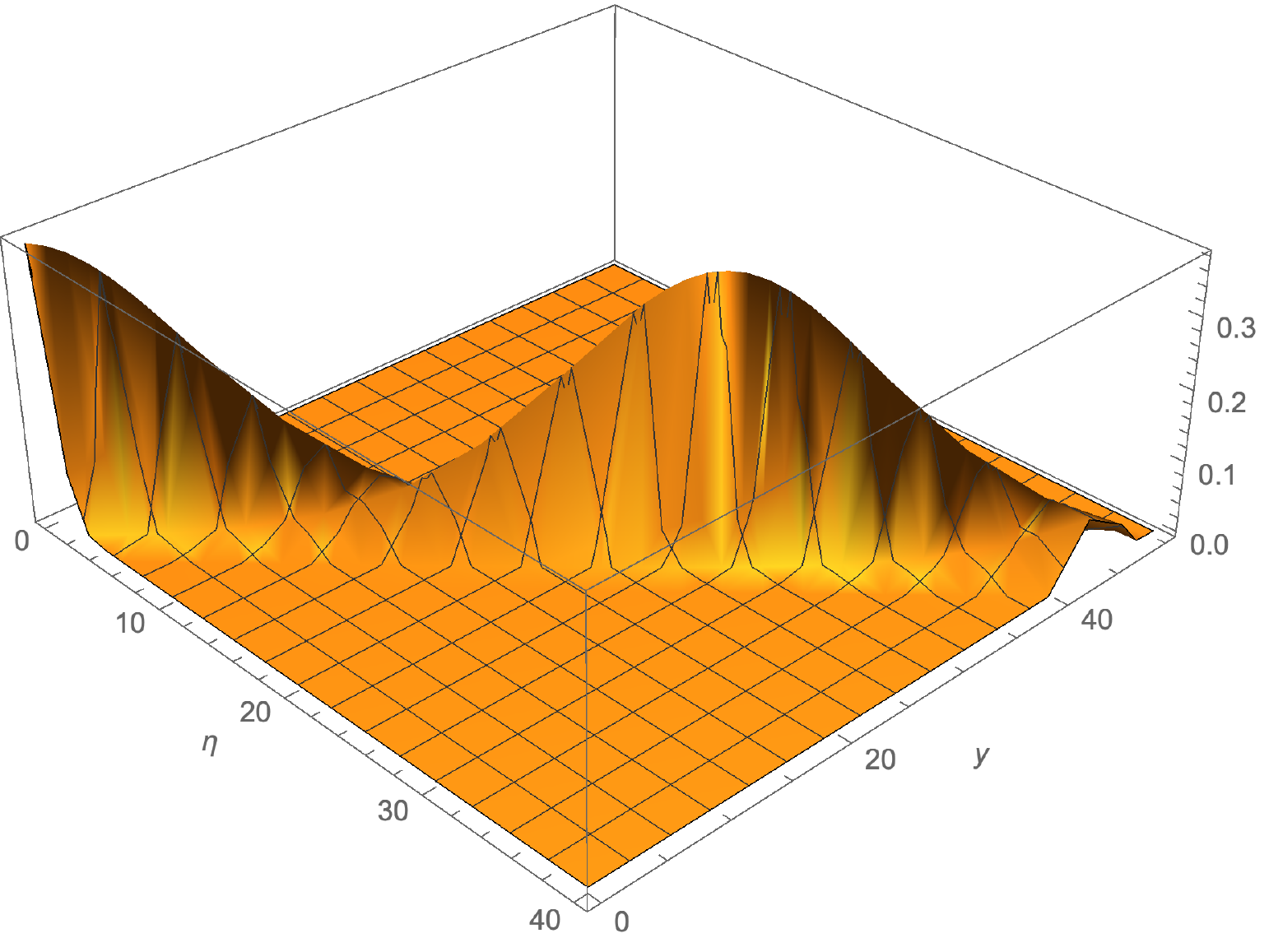}\quad
\includegraphics[width=0.45\linewidth]{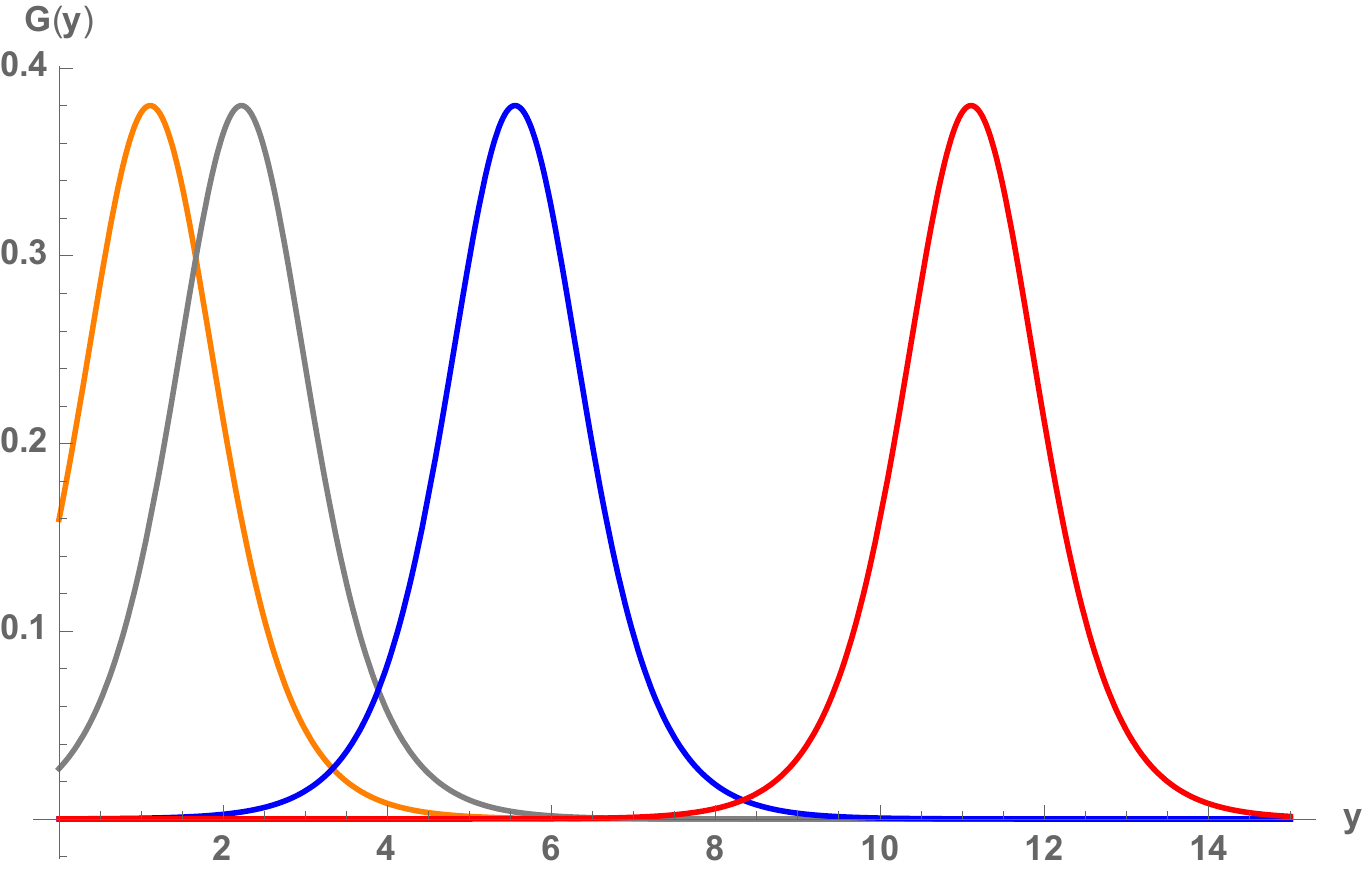}
\caption{Left: 3D Plot of $\fn{G}{\eta,y}$ with $\Lambda=c_1=1, \,c_2=-0.9$, and  $c_3=0$. {Right}: 1D snapshots of the same $\fn{G}{\eta,y}$ at $\eta={ 1},  {2}, { 5},  { 10}$ from left to right. The `soliton' maintains its shape in space and propagates in time. $\Lambda=1$ units are used.}
\label{fig:conformal}
\end{figure}

\begin{figure}[H]
\includegraphics[width=0.6\linewidth]{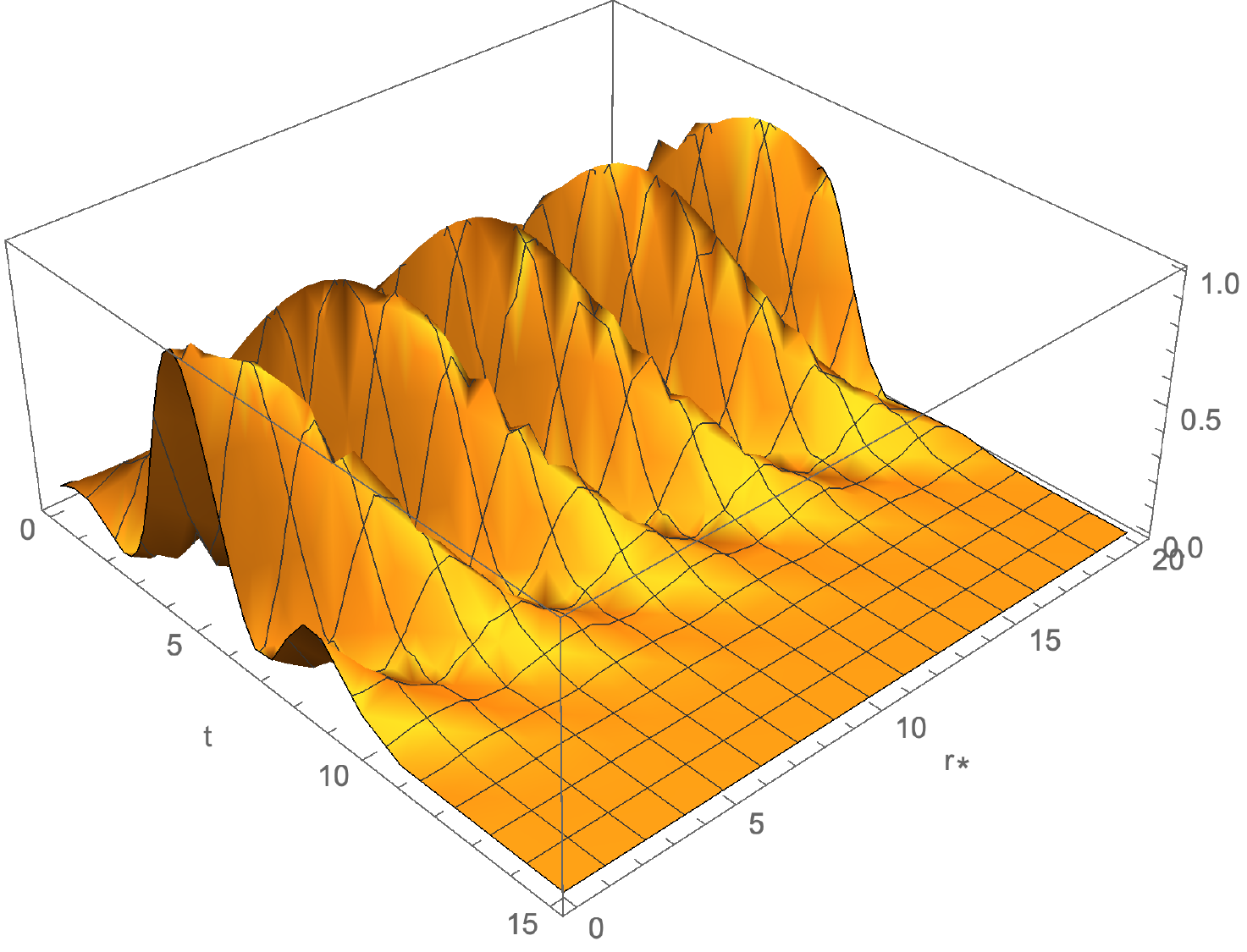}
\caption{3D Plot of $\fn{H}{\tilde t,r_*}=e^{-0.1(\tilde t-5)^2}\cos\left[ \sqrt{\frac{\Lambda}{2}}\left( r_*-\tilde t \right)\right]$ with $\Lambda=1$. The wavepacket dies out quickly with time. $\Lambda=1$ units are used.}
\label{fig:spatiallyflat}
\end{figure}

Let us explicitly show that we can move from one coordinate choice to another.
 The first observation is that the background solution $p(r)$, can be easily transformed into the spatially flat gauge by defining the tortoise coordinate  $dr_*=\frac{dr}{\sqrt{\fn{p}{r}}}$. Similarly, we can transform our ansatz to the conformally flat frame as follows. Consider first the background solution, $p(r)$:
\be
ds^2=p(r)\left(-dt^2+\frac{dr^2}{p(r)^2}\right)\equiv p(y)(-d\eta^2+dy^2),
\ee
where we have used $\eta\equiv t$ and again a tortoise coordinate, but this time with $dy=\frac{dr}{p(r)}$, giving $y=-\frac{2 \tanh^{-1} \left(\frac{-B+\Lambda r}{\sqrt{B^2+2 A \Lambda}}\right)}{\sqrt{B^2+2A\Lambda}}$. One can also write down $p(y)$ explicitly:
\be
p(y)=\frac{B^2+2A \Lambda}{\Lambda\left[1+\cosh(\sqrt{B^2+2 A \Lambda} y)\right]}.
\ee

Hence we have achieved an explicit transformation between the background solution in Schwarzschild-like  coordinates to the spatially flat and conformally flat gauge. Notice that in the conformally flat case the ``horizons'' are pushed to infinity.
Now let us look at our traceless perturbation. The perturbed metric can now be written as:
\begin{equation} 
g_{\mu \nu} = 
\begin{pmatrix} 
-\fn{p}{y}(1-\fn{\tilde h}{\tau,y}) & 0 \\ 
0 & \fn{p}{y}(1+ \fn{\tilde h}{\tau,y} )
\end{pmatrix} 
\end{equation}

\textbf{First Order: 
}

Let us substitute $\tilde h(\tau,y)\propto e^{-i \omega \tau}Y(y)$. Using the zeroth order results yields:
\be
Y''+\frac{p'}{p}Y'-\left(\Lambda p+\omega^2\right)Y=0,
\ee
with the following general solution:
\be
Y(y)=\left(c_1 e^{-\frac{y}{2}\sqrt{4\omega^2+B^2+2A\Lambda}}+c_2e^{\frac{y}{2}\sqrt{4\omega^2+B^2+2A\Lambda}}\right)\cosh\left(\sqrt{B^2+2A\Lambda}\frac{y}{2}\right)
\ee

Notice that now we can get all kinds of behavior depending on the sign of the various parameters and specifically $\Lambda$.
For instance, a positive CC will generally give an exponential solution. However, $\Lambda<0$ can give an oscillating solution.
Consider $4\omega^2+B^2+2A\Lambda<0$ which is possible for a negative enough $\Lambda<0$ and a small enough frequency $\omega$. Defining $4\omega^2+B^2+2A\Lambda=-\gamma^2$ with suitable initial conditions gives the following result:
\be
Y(y)=\cos\left(\frac{\gamma}{2}y\right)\cos\left(\frac{|B^2+2A\Lambda|}{2}y\right)
\ee

\textbf{Second order:}

Considering the backreaction on the metric for $\langle (e^{-i \omega t})^2\rangle =1/2$ gives the following equation:
\be
\tilde p^3 \Lambda=(\tilde p'^2-\tilde p \tilde p'')\left(1+\frac{Y^2}{2}\right)-\frac{1}{2}Y Y' \tilde p \tilde p'
\ee

This has to be solved numerically, with the exception of $\Lambda=0$. In such a case, one has an exact solution
\be
\tilde p(y)=c_1e^{c_2\int^y\frac{du}{\sqrt{1+Y(u)^2/2} }}.
\ee

Hence, we have demonstrated that our analysis is not limited to a specific gauge coordinate system, and can be done by choosing the most convenient coordinate system and gauge. {In all gauges, the perturbation cannot be removed by a coordinate or gauge transformation, and backreacts on the geometry. Therefore, it is a true physical entity and not an artefact.}

\section{Discussion}\label{dis}

In this note, we have discussed the possibility of GW trapped in space within the context of two-dimensional Jackiw-Teitelboim gravity. 
We have shown the existence of a vacuum solution for such gravitational waves numerically and analytically and tested the stability of the solution against its backreaction on the background metric. This is expected on general grounds as the gravitational clump experiences dispersion on the one hand and gravitational attraction on the other hand. The CC further induces a potential term. As a consequence, the clump sits at the minimum of a gravitational potential which is attractive in the vicinity of the clump and repulsive at large scales. Our gauge invariance analysis established that the theory has a single physical degree of freedom, and we showed how to obtain trapped solutions in different coordinate systems and different gauges. The clearest results were achieved in the traceful gauge.
In this simple example, we summarized two guiding principles of trapping the gravitational waves in space by inspecting the asymptotic behaviors, AB1 and AB2: {The trace waves 
 would be trapped in the region where the metric component $p(r)$ goes to zero but not in the region where the it becomes divergent.}
In other gauges, the principle is less clear, but we have shown that a single degree of freedom always exists and backreacts on the metric. Specifically, the off-diagonal choice gives very simple solutions, even at second order and allows for exact analytical treatment. 
 
Works \cite{Brill:1964zz, Anderson:1996pu},  already suggested the possibility that a gravitational geon can be attained. Although our solution is nothing but a toy model,  it represents a new approach to the self-gravitating gravitational perturbations in an expanding Universe. The notion of the active region where all the fluctuation's effective energy is deposited---an approach that raises doubts about stability of the configuration---is abandoned.  Instead, our solution relinquishes the effective region and stability is achieved by competing terms in the gravitational potential due to the cosmological evolution.

The same general arguments discussed here are expected to hold in our expanding Universe, since if we take a spherically symmetric perturbation it will again feel the competing effect of its gravitational attraction and the CC. Our metric ansatz corresponds to spaces with `horizons,' which are found at the points $\fn{p}{r}=0$ in our representation.  Should we obtain a solution, that resembles the 2D obtained in this article (as in \mbox{Figures \ref{fig:L=0} and \ref{fig:L0.2}}), we conjecture then that these perturbations (which are truly gravitational waves in 4D)  may be trapped behind a ``black hole'' horizon and hence, unobservable. Nevertheless, they could leave traces of such trapping in the form of diffused remnants. However, by far the most promising example would be that of Figure \ref{fig:standing_desitter} that corresponds to a modified Schwarzschild--de Sitter space, according to \cite{Sikkema:1989ib}.  In this case, the gravitational waves are similarly  trapped between the ``black hole'' and the cosmological  horizon in the expanding dS-like space. Clearly, the stability should depend on the rate of expansion of the Universe: a slowly expanding Universe does not provide enough repulsion for stability---such a clumped  solution should be meta-stable. 
A current work in progress for confronting the 4D case, is by considering Schwarzschild-de Sitter space in static coordinates with GW perturbations. Expanding these in spherical harmonics and integrating over the angular variables will yield a 2D effective theory similar to the one we have analyzed, \cite{Martinon:2017esl}. We further speculate that such geons may have formed in the early universe when expansion was sufficiently fast, existed for some time, and got dispersed as the Universe expanded. If that is the case, it might be that the fingerprints of these objects can be observable in the form of a diffused spectrum of GWs  that have propagated and undergone redshift.

\section*{Acknowledgements}
The authors thank to Dong-han Yeom and Sungwook E. Hong. This work is supported by the DGIST  UGRP grant.

\bibliographystyle{apsrev4-1}
\bibliography{geon}

\end{document}